\documentclass[twocolumn]{aastex631}

\usepackage{multirow}
\usepackage{hyperref}  

\usepackage{amsmath}

\defcitealias{Calzetti+00}{C00}
\defcitealias{Bruzual+03}{BC03}

\begin{document}

\title{MIDIS: MIRI uncovers {\it Virgil}, the first Little Red Dot with clear detection of its host galaxy at $z\simeq 6.6$}

\correspondingauthor{Edoardo Iani}
\email{iani@astro.rug.nl, E.Iani@rug.nl}

\author[0000-0001-8386-3546]{Edoardo Iani}
\affiliation{Kapteyn Astronomical Institute, University of Groningen, P.O. Box 800, 9700 AV Groningen, The Netherlands}
\affiliation{Institute of Science and Technology Austria (ISTA), Am Campus 1, 3400 Klosterneuburg, Austria}

\author[0000-0002-5104-8245]{Pierluigi Rinaldi}
\affiliation{Steward Observatory, University of Arizona, 933 North Cherry Avenue, Tucson, AZ 85721, USA}
\affiliation{Kapteyn Astronomical Institute, University of Groningen, P.O. Box 800, 9700 AV Groningen, The Netherlands}

\author[0000-0001-8183-1460]{Karina I. Caputi}
\affiliation{Kapteyn Astronomical Institute, University of Groningen, P.O. Box 800, 9700 AV Groningen, The Netherlands}
\affiliation{Cosmic Dawn Center (DAWN), Denmark}

\author[0000-0002-8053-8040]{Marianna Annunziatella}
\affiliation{Centro de Astrobiolog\'ia (CAB), CSIC-INTA, Ctra. de Ajalvir km 4, Torrej\'on de Ardoz, E-28850, Madrid, Spain}

\author[0000-0001-5710-8395]{Danial Langeroodi}
\affiliation{DARK, Niels Bohr Institute, University of Copenhagen, Jagtvej 155, 2200 Copenhagen, Denmark}

\author[0000-0003-0470-8754]{Jens Melinder}
\affiliation{Department of Astronomy, Stockholm University, Oscar Klein Centre, AlbaNova University Centre, 106 91 Stockholm, Sweden}

\author[0000-0003-4528-5639]{Pablo G. P\'erez-Gonz\'alez}
\affiliation{Centro de Astrobiolog\'ia (CAB), CSIC-INTA, Ctra. de Ajalvir km 4, Torrej\'on de Ardoz, E-28850, Madrid, Spain}

\author[0000-0002-7093-1877]{Javier \'Alvarez-M\'arquez}
\affiliation{Centro de Astrobiolog\'ia (CAB), CSIC-INTA, Ctra. de Ajalvir km 4, Torrej\'on de Ardoz, E-28850, Madrid, Spain}

\author[0000-0002-3952-8588]{Leindert A. Boogaard}
\affiliation{Max Planck Institute for Astronomy, K\"onigstuhl 17, 69117 Heidelberg, Germany}

\author[0000-0001-8582-7012]{Sarah E. I. Bosman}
\affiliation{Max Planck Institute for Astronomy, K\"onigstuhl 17, 69117 Heidelberg, Germany}
\affiliation{Institute for Theoretical Physics, Heidelberg University, Philosophenweg 12, D–69120, Heidelberg, Germany}

\author[0000-0001-6820-0015]{Luca Costantin}
\affiliation{Centro de Astrobiolog\'ia (CAB), CSIC-INTA, Ctra. de Ajalvir km 4, Torrej\'on de Ardoz, E-28850, Madrid, Spain}

\author[0000-0002-3305-9901]{Thibaud Moutard}
\affiliation{European Space Agency (ESA), European Space Astronomy Centre (ESAC), Camino Bajo del Castillo s/n, 28692 Villanueva de la
Cañada, Madrid, Spain}

\author[0000-0002-9090-4227]{Luis Colina}
\affiliation{Centro de Astrobiolog\'ia (CAB), CSIC-INTA, Ctra. de Ajalvir km 4, Torrej\'on de Ardoz, E-28850, Madrid, Spain}

\author[0000-0002-3005-1349]{G\"oran \"Ostlin}
\affiliation{Department of Astronomy, Stockholm University, Oscar Klein Centre, AlbaNova University Centre, 106 91 Stockholm, Sweden}

\author[0000-0002-2554-1837]{Thomas R. Greve}
\affiliation{Cosmic Dawn Center (DAWN), Denmark}
\affiliation{DTU Space, Technical University of Denmark, Elektrovej, Building 328, 2800, Kgs. Lyngby, Denmark}
\affiliation{Department of Physics and Astronomy, University College London, Gower Street, WC1E 6BT, UK}

\author[0000-0001-7416-7936]{Gillian Wright}
\affiliation{UK Astronomy Technology Centre, Royal Observatory Edinburgh, Blackford Hill, Edinburgh EH9 3HJ, UK}

\author[0000-0001-6794-2519]{Almudena Alonso-Herrero}
\affiliation{Centro de Astrobiolog\'ia (CAB), CSIC-INTA, Camino Bajo del Castillo s/n, E-28692 Villanueva de la Ca\~nada, Madrid, Spain}

\author[0000-0001-8068-0891]{Arjan Bik}
\affiliation{Department of Astronomy, Stockholm University, Oscar Klein Centre, AlbaNova University Centre, 106 91 Stockholm, Sweden}

\author[0000-0001-9885-4589]{Steven Gillman}
\affiliation{Cosmic Dawn Center (DAWN), Denmark}
\affiliation{DTU Space, Technical University of Denmark, Elektrovej, Building 328, 2800, Kgs. Lyngby, Denmark}

\author[0000-0003-2119-277X]{Alejandro Crespo G\'omez}
\affiliation{Centro de Astrobiolog\'ia (CAB), CSIC-INTA, Ctra. de Ajalvir km 4, Torrej\'on de Ardoz, E-28850, Madrid, Spain}

\author[0000-0002-4571-2306]{Jens Hjorth}
\affiliation{DARK, Niels Bohr Institute, University of Copenhagen, Jagtvej 128, 2200 Copenhagen, Denmark}

\author[0000-0002-7612-0469]{Sarah Kendrew}
\affiliation{European Space Agency (ESA), ESA Office, Space Telescope Science Institute, 3700 San Martin Drive, Baltimore, MD 21218, USA}

\author[0000-0002-0690-8824]{Alvaro Labiano}
\affiliation{Telespazio UK for the European Space Agency, ESAC, Camino Bajo del Castillo s/n, E-28692 Villanueva de la Ca\~nada, Madrid, Spain}
\affiliation{Centro de Astrobiolog\'ia (CAB), CSIC-INTA, Camino Bajo del Castillo s/n, E-28692 Villanueva de la Ca\~nada, Madrid, Spain}

\author[0000-0002-0932-4330]{John P. Pye}
\affiliation{School of Physics \& Astronomy, Space Park Leicester, University of Leicester, 92 Corporation Road, Leicester LE4 5SP, UK}

\author[0009-0003-6128-2347]{Tuomo V. Tikkanen}
\affiliation{School of Physics \& Astronomy, Space Park Leicester, University of Leicester, 92 Corporation Road, Leicester LE4 5SP, UK}

\author[0000-0003-4793-7880]{Fabian Walter}
\affiliation{Max-Planck-Institut f\"ur Astronomie, K\"onigstuhl 17, 69117 Heidelberg, Germany}

\author[0000-0001-5434-5942]{Paul P. van der Werf}
\affiliation{Leiden Observatory, Leiden University, P.O. Box 9513, 2300 RA Leiden, The Netherlands}

\begin{abstract}

We present {\it Virgil}, a MIRI extremely red object (MERO) detected with the F1000W filter as part of the MIRI Deep Imaging Survey (MIDIS) observations of the Hubble Ultra Deep Field (HUDF). 
{\it Virgil} is a Lyman-$\alpha$ emitter (LAE) 
at $z_{spec} = 6.6312\pm 0.0019$ (from VLT/MUSE) with a rest-frame UV-to-optical spectral energy distribution (SED) typical of LAEs at similar redshifts.
However, MIRI observations reveal an unexpected extremely red color at rest-frame near-infrared wavelengths, $\rm F444W - F1000W = 2.33 \pm 0.06$.
Such steep rise in the near-infrared, completely missed without MIRI imaging, is poorly reproduced by models including only stellar populations and hints towards the presence of an Active Galactic Nucleus (AGN).
According to the shape of its overall SED, {\it Virgil} belongs to the recently discovered population of Little Red Dots (LRDs) but displays an extended rest-frame UV-optical wavelengths morphology 
following a 2D-S\'ersic profile with average index $n = 0.93^{+0.85}_{-0.31}$ and $r_e = 0.49^{+0.05}_{-0.11}$~pkpc. 
Only at MIRI wavelengths {\it Virgil} is unresolved due to the coarser PSF.
We also estimate a bolometric luminosity $L_{\rm AGN, bol} = (8.9 - 11)\times 10^{44}\rm~erg~s^{-1}$ and a supermassive black hole mass $M_{\rm BH} = (7 - 9)\times 10^6\rm ~ M_\odot$ (if $\lambda_{\rm Edd} = 1$) in agreement with recently reported values for LRDs.
This discovery demonstrates the crucial importance of deep MIRI surveys to reveal the true nature and properties of high-$z$ galaxies that otherwise would be misinterpreted and raises the question of how common {\it Virgil}-like objects could be in the early Universe. 
\end{abstract}

\keywords{Galaxy formation (595) --- Galaxy evolution (594) --- High-redshift galaxies (734) --- Stellar populations (1622) --- Broad band photometry (184) --- Galaxy ages (576) --- JWST (2291) --- Active galactic nuclei (16)}

\section{Introduction}

Since the release of its first observations in July 2022, JWST has been opening up a new window on studying the early Universe \citep{Gardner+23}, both by giving us exceptional details on known high-z sources 
\citep[e.g., ][]{Bunker+23, Maiolino+23, Colina+23} and revealing the presence of new galaxies
\citep[e.g., ][]{Atek+23a, Atek+23b, Casey+23a, Robertson+23}, some of which are high-z ($z>10$) candidates \citep[e.g., ][]{Iani+22, Rodighiero+23, Gandolfi+25} while others are spectroscopically confirmed (e.g., \citealt{Curtis_Lake+23, Bunker+23, Wang+23, Castellano+24}).

Thanks to JWST new galaxy populations have also been discovered, e.g. galaxies with very high equivalent width 
emission lines \citep{Rinaldi+23a, Boyett+24, vanMierlo+23, Alvarez-Marquez+24} and the enigmatic class of the so-called “Little Red Dots” \cite[LRDs; e.g., ][]{Labbe+23b, Matthee+24, Barro+23, Kokorev+24, Langeroodi2023c, Perez-Gonzalez+24a}, i.e. compact sources with distinct spectral energy distributions showing clear Lyman and Balmer breaks and a characteristic ``v-shaped'' SED (in the $\lambda$ -- $f_\lambda$ plane) implying blue rest-frame UV-to-optical colors and red optical-to-infrared colors. Their nature is still under strong debate \cite[e.g., ][]{Labbe+23b, Li+24, Inayoshi+24, Bellovary+25}.

In this context, the Mid-Infrared Instrument \cite[MIRI;][]{Rieke+15, Wright+15, Wright+23} on JWST and, in particular, its imager 
\cite[MIRIM, ][]{Bouchet+15, Dicken+24} is allowing us to unfold the intermediate/high-redshift 
Universe by enabling us to study galaxies in the wavelength range from $5.6\;\mu$m 
through $25.5\;\mu$m at an unprecedented spatial resolution and sensitivity \citep{Rigby+23}. 
Among its numerous results, MIRIM has been crucial in revealing previously undetected faint galaxies \cite[e.g., ][]{Kirckpatrick+23, Barro+23, Akins+23, Perez-Gonzalez+24a}, 
uncovering a large population of obscured AGNs \cite[e.g., ][]{Yang+23}, and characterising the physical properties of distant 
galaxies in great detail \cite[e.g., ][]{Colina+23, Alvarez-Marquez+23}, including their morphologies \cite[e.g., ][]{Boogaard+23, Costantin+24, Gillman+24}. 
MIRI also has a unique role in the study of early galaxies. 
By leveraging the deepest image of the Universe at $5.6~\mu$m 
\citep{Ostlin+25}, \citet{Rinaldi+23a} 
inferred for the first time the presence of H$\alpha$ emission in individual galaxies at $z\approx 7-9$ via broad-band photometric excess.

Thanks to its capabilities, MIRIM holds the promise to reveal and constrain the rest-frame optical-NIR properties of galaxies at $z > 2$ and further push for the discovery and study of the class of ``Extremely Red Objects'' \citep[EROs; e.g.,][]{Elston+88, Graham+96, Caputi+04}, i.e. sources displaying a large (red) color between bands probing their rest-frame optical and near-infrared emission \cite[e.g. R - K $>$ 5 in the Vega system, ][]{Elston+88}. 
In this regard, particularly remarkable is the recent discovery of the first ERO detected thanks to MIRI imaging at 10~$\mu$m by \cite{Perez-Gonzalez+24b}: with a F444W - F1000W $>$ 3.5 color, this object represented the first MIRI Extremely Red Object (MERO) reported in the era of JWST.


In this paper, we present and study {\it Virgil}, a MERO found in the MIRI Deep Imaging Survey \cite[MIDIS, ][]{Ostlin+25} F1000W imaging ($\alpha$(J2000.0) = 03:32:37.9370 (hours); $\delta$(J2000.0) = -27:47:10.712 (degrees)). 
The source is clearly detected in the NIRCam bands and it has been already catalogued in both JADES DR2 and VLT/MUSE catalogs \citep{Bacon+23, Rieke+23}. 
{\it Virgil} is a Lyman-$\alpha$ emitter (LAE) at $z_{spec} = 6.6312 \pm 0.0019$. 
that shows an extremely red color between 4.4 and 10 $\mu$m and is also red with respect to the bluest MIRI band (F560W), as revealed by MIDIS F560W imaging \citep{Ostlin+25}.
Without the MIRIM coverage, this unexpected property of {\it Virgil} would have been completely missed, thus leaving us with a partial and biased knowledge of it. 

This paper is organized as follows. In \S\ref{sec:data} we briefly describe the datasets used and in particular the MIDIS 10~$\mu$m MIRI imaging and the ancillary HST and JWST NIRCam and MIRI observations.
All the HST and JWST observations can be found in the Mikulski Archive for Space Telescopes (MAST\footnote{\url{https://mast.stsci.edu/portal/Mashup/Clients/Mast/Portal.html}})\footnote{The HST and JWST observations can be found at the following DOIs: 
\href{https://doi.org/10.17909/t91019}{10.17909/t91019},
\href{https://doi.org/10.17909/fsc4-dt61}{10.17909/fsc4-dt61}, 
\href{https://doi.org/10.17909/gdyc-7g80}{10.17909/gdyc-7g80}, 
\href{https://doi.org/10.17909/8tdj-8n28}{10.17909/8tdj-8n28}, 
\href{https://doi.org/10.17909/et3f-zd57}{10.17909/et3f-zd57}, 
\href{https://doi.org/10.17909/5txh-pj89}{10.17909/5txh-pj89},  \href{https://doi.org/10.17909/bjk7-qh92}{10.17909/bjk7-qh92}.}.
In \S\ref{sec:virgil} we describe the identification of {\it Virgil}, the assessment of its redshift, the decontamination of its light from the nearby LAE at $z\simeq 4.77$, the analysis of its morphology, and the extraction of its multi-wavelength photometry.
In \S\ref{sec:discussion} we discuss the possible nature and the importance of MIRI imaging for the detection and characterisation of such an object. 
Finally, we summarise our findings and present our conclusions in \S\ref{sec:conclusions}.

Throughout the paper, we assume a flat $\Lambda$-CDM cosmology with $\Omega_{M} = 0.3$, $\Omega_{\Lambda} = 0.7$, and a Hubble constant 
$\rm H_{0} = 70\;km\,s^{-1}\,Mpc^{-1}$. 
According to this cosmology, the luminosity distance of our target is $D_{\rm L} = 64.8$~Gpc and the age of the universe at its redshift was about 0.8 Gyr.
We adopt the AB magnitudes \citep{Oke+83}. 
All stellar mass and SFR estimations assume a universal Chabrier initial mass function \citep[IMF;][]{Chabrier+03}.
Finally, unless differently stated, we report the average 5$\sigma$ depth for point-like sources of the available NIRCam and MIRI imaging of the MIDIS area as measured in circular apertures of r = 0\farcs2 (NIRCam) and r = 0\farcs3 (MIRI) as outlined by 
\citet{Ostlin+25} and taking into account effects of pixel correlation as explained in \cite{Fruchter+02}.
The depth estimates are not corrected for aperture.


\section{Datasets}
\label{sec:data}
\subsection{MIRI F1000W data}

MIRI data in the F1000W filter were taken as part of {\it The MIRI Deep Imaging Survey} (MIDIS, PID: 1283; PI: G. \"Ostlin; \citealt{Ostlin+25}) in December 2023. 
The observation consisted of 11 exposures, each with 100 groups and 7 integrations, for a total on-source exposure time of 30.8~ks, 
centered on the Hubble Ultra Deep Field \cite[HUDF,][]{Beckwith+06}. 
The dithering pattern was set to large-size cycling, with the 11 exposures taken in different positions on the sky separated by up to 10\arcsec. 
This dithering pattern, with no repeated positions, was selected since it was found to be crucial for the detection of faint sources, such as {\it Virgil}, and their distinction from detector artifacts. 

\begin{figure*}
    \centering
    \includegraphics[width = .8\textwidth]{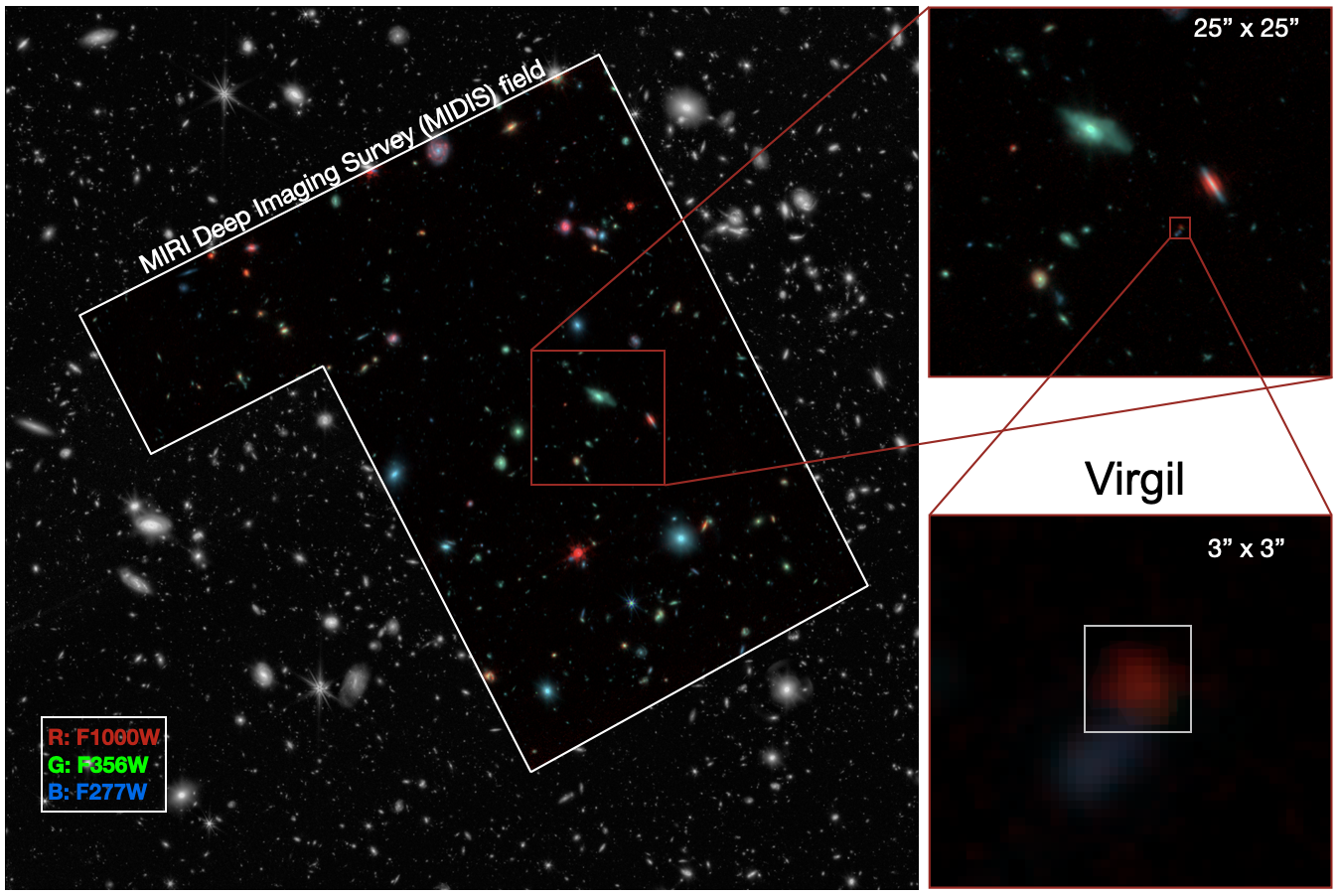}\\
    \includegraphics[width = \textwidth]{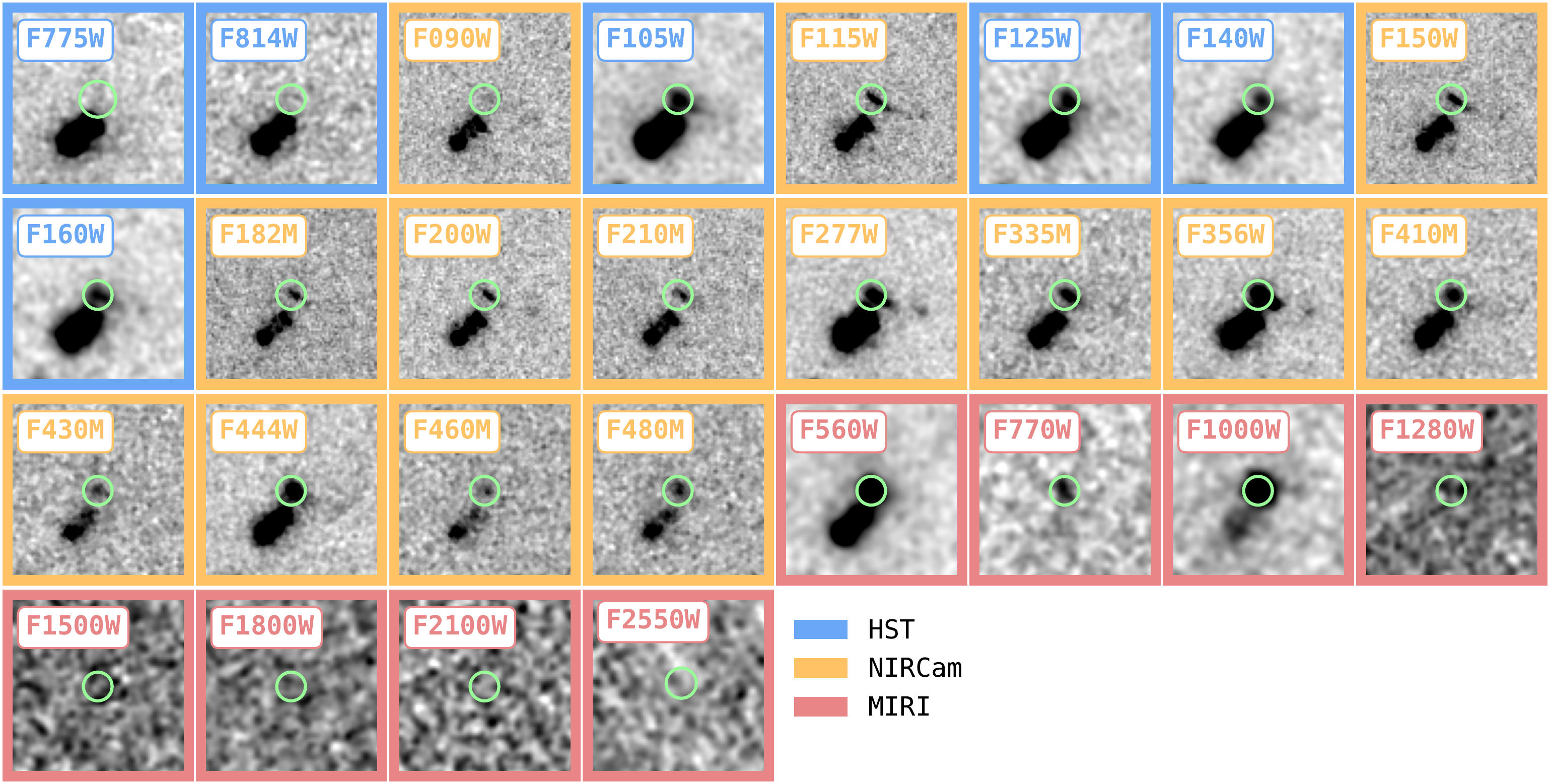}
    \caption{{\bf Top Panel:} RGB composition of the MIDIS field. 
    The color images have been built with JADES data in 2 NIRCam filters, F277W and F356W, and the MIDIS MIRI F1000W filter (all convolved to the same PSF as MIRI/F1000W). In the background, we show the HUDF JADES data in gray scale. 
    We show a series of zoomed-in RGB frames that lead to the MIRI Extremely Red Object source, {\it Virgil} ($\alpha$(J2000.0) = 03:32:37.9370 (hours); $\delta$(J2000.0) = -27:47:10.712 (degrees)). The {\it Virgil} object is highlighted with a square in the RGB frame at the bottom right.
    {\bf Bottom Panel:} Cutouts (2\farcs5 $\times$ 2\farcs5) of {\it Virgil} from HST/ACS and WFC3 (F775W, F814W, F105W, F125W, F140W, and F160W), JWST/NIRCam 
    (F090W, F115W, F150W, F200W, F210M, F277W, F335M, F356W, F410M, F430M, F444W, F460M, and F480M) and JWST/MIRI 
    (F560W, F770W, F1000W, F1280W, F1500W, F1800W, F2100W, and F2550W). 
    We highlight {\it Virgil} with a green circle ($\rm r = 0\farcs2$.) 
    HST cutouts below 0.7 $\mu$m are not shown here.}
    \label{fig:midis}
\end{figure*}

A detailed description of the reduction of the MIRI data is presented in \citet{Perez-Gonzalez+24b}. 
Briefly, a super-background strategy is used to build background maps for every single image using all the other exposures 
(since they were taken during the same campaign), which results in a very homogeneous background (in terms of level and noise). 
Known sources are masked to avoid biasing the determination of the very local background in the super-background frame. 
Our final F1000W image, reduced with a 60 milliarcsec pixel-scale, presents an average 5$\sigma$ depth of 26.4~mag. 
We present the MIDIS F1000W data in Figure~\ref{fig:midis} (top panels) as the red channel of an RGB image built in combination with the NIRCam long-wavelength channels F277W and F356W.

\subsection{Ancillary data}

We complement our MIRI observations at 10~$\mu$m with a rich set of photometric and spectroscopic data publicly available. 
In particular, we extend our dataset including more MIRI observations, HST and JWST/NIRCam filters, 
the Atacama Large Millimeter/submillimeter Array (ALMA) surveys and deep optical integral field spectroscopy by the 
Multi Unit Spectroscopic Explorer (MUSE) at the Very Large Telescope (VLT). We briefly summarise these ancillary datasets in the following sections. 

\subsubsection{Additional MIRI data}

We combine the recently acquired MIDIS F1000W data with the F560W ultra-deep observations also carried out by MIDIS in December 2022 and January 2023. 
A thorough description of this dataset is presented in \citet{Ostlin+25}. 
These observations consist of $\sim41$~hours on source taken in the HUDF, reaching a 5$\sigma$ depth of 28.6~mag.
We add to the MIDIS observations the publicly available MIRI data gathered by the 
{\it Systematic Mid-infrared Instrument Legacy Extragalactic Survey} \cite[SMILES, PID 1207; PI: G. Rieke;][]{Lyu+23, Rieke+24, Alberts+24} in the 
F770W, F1280W, F1500W, F1800W, F2100W and F2550W bands.
We reduce the SMILES images following the same methodology applied to the MIDIS dataset and obtain images with a 5$\sigma$ depth 
of about 26.2 (F770W), 25.0 (F1280W), 24.9 (F1500W), 24.1 (F1800W), 23.6 (F2100W) and 21.8 (F2550W) mag, respectively.

\subsubsection{NIRCam data}

We complement the MIRI data with NIRCam imaging taken by {\it The JWST Advanced Deep Extragalactic Survey} 
\citep[JADES, PIDs: 1180, 1210; P.I.: D. Eisenstein, N. Luetzgendorf;][]{2023arXiv230602465E}, Data Release 2  
(\citealt{2023arXiv231012340E}), which includes also observations from the {\it JWST Extragalactic Medium-band Survey} 
\cite[JEMS, PID: 1963; PIs: C. C. Williams, S. Tacchella, M. Maseda; ][]{2023ApJS..268...64W} and 
{\it The First Reionization Epoch Spectroscopically Complete Observations} \cite[FRESCO, PID: 1895; PI: P. Oesch;][]{2023MNRAS.525.2864O}. 
This dataset provides a total of 14 bands from 0.9 to 4.8~$\mu$m (F090W, F115W, F150W, F182M, F200W, F210M, F277W, F335M, F356W, F410M, F430M, F444W, F460M, F480M) 
with 5$\sigma$ depths ranging from 30.5 to 30.9~mag.

\subsubsection{HST data}

We obtain all the HST images over the HUDF from the Hubble Legacy Field GOODS-S \cite[HLF-GOODS-S][]{Whitaker+19}. 
The HLF-GOODS-S provides 13 HST bands covering a wide range of wavelengths (0.2$-$1.6$\mu$m), 
from the UV (WFC3/ UVIS F225W, F275W, and F336W filters), optical (ACS/ WFC F435W, F606W, F775W, F814W, and F850LP filters), 
to near-infrared (WFC3/IR F098M, F105W, F125W, F140W and F160W filters). 
In this paper, we only made use of the deepest ones (i.e., F435W, F606W, F775W, F814W, F850LP, F105W, F125W, F140W, F160W). 
We refer the reader to \citet{Whitaker+19} for more detailed information on these observations.\footnote{The HLF-GOODS-S imaging is available at \url{https://archive.stsci.edu/ prepds/hlf/};}

\subsubsection{MUSE data}

The HUDF has been extensively studied with MUSE \citep{2010SPIE.7735E..08B} at VLT as part of the MOSAIC and UDF-10 fields 
(GTO programs 094.A- 0289(B), 095.A-0010(A), 096.A-0045(A) and 096.A- 0045(B), PI: R. Bacon) and the most recent MXDF observations 
(GTO Large Program 1101.A-0127, PI: R. Bacon). 
These observations cover a spectral wavelength range between 4750 - 9350\,{\AA} with a spectral resolving power (R) that varies 
from 1770 (4800\,{\AA}) to 3590 (9300\,{\AA}) and have a PSF (at 7750 \AA), that can be described by a Moffat function \citep{Moffat+69} 
that varies from FWHM = 0\farcs45 and $\beta = 1.89$ (MXDF), to FWHM = 0\farcs60 and $\beta = 2.80$ (UDF-10), 
and to FWHM = 0\farcs63 and $\beta$ = 2.80 (MOSAIC), see \cite{Bacon+23} and their Table 4.
In the MIDIS area, these observations have exceeded total exposure times of 140 hours, although the 
distribution of the depth in flux is not uniform across the whole area.
More details about the HUDF MUSE surveys can be found in \citet{Bacon+17, 2023A&A...670A...4B}. In our study we use the fully reduced MUSE datacubes for the MOSAIC + UDF-10 and MXDF programs, 
as well as the catalog of detected sources and corresponding redshifts presented in \citet{Bacon+23}.\footnote{The MUSE cubes can be obtained at \url{https://amused.univ- lyon1.fr/project/UDF/HUDF/};}

\subsubsection{ALMA data}

The Atacama Large Millimeter/submillimeter Array (ALMA) Spectroscopic Survey in the Hubble Ultra Deep Field (ASPECS) is a 
Cycle 4 Large Program over a 4.6 arcmin$^{2}$ scan at 1.2~mm \cite[band 6; ][]{Decarli+20, gonzalez-lopez2020} and 
3.0~mm \cite[band 3;][]{Decarli+19, gonzalez-lopez2019}. 
The ultra-deep 1.2\,mm data reaches an rms sensitivity of 9.3\,$\mu$Jy/beam, with beam dimensions of $1\farcs5\times1\farcs1$.  
The 3.0\,mm data reaches 1.4\,$\mu$Jy/beam, with a $1\farcs8\times1\farcs5$ beam.
We download the fully reduced ALMA images from the official ASPECS website.\footnote{The ASPECS images are available at \url{https://almascience.org/alma-data/lp/ASPECS};}


\section{Selection and properties of the MIRI extremely red Object {\it Virgil}}
\label{sec:virgil}
In this paper, we present {\it Virgil} ($\alpha$(J2000.0) = 03:32:37.9370 (hours); $\delta$(J2000.0) = -27:47:10.712 (degrees)), a source in the deep MIDIS F1000W image of the HUDF, which is counterpart to a LAE at $z=6.6312$ from MUSE/VLT (\S\ref{sec:redshift}).
Despite being already known from shorter wavelength data \citep{Bacon+23, 2023ApJS..269...16R,2023arXiv231012340E}, new important insights on the nature of this object come from the new deep MIRI observations 
at 10~$\mu$m. 
In particular, {\it Virgil} displays an extreme F444W $-$ F1000W red color ($\geq 2$) that classifies it as an ERO. 
Since the ERO nature of this source can only be determined with MIRI, we refer to it as MERO. 

\subsection{Identification of the MERO {\it Virgil}}
Motivated by the recently reported discovery of the MERO {\it Cerberus} found in the MIDIS field \cite[$\rm F444W - F1000W > 3.5$;][]{Perez-Gonzalez+24b}, we decide to further scout the MIDIS area to discover additional MIRI red objects.
To do so, we make use of the \textsc{sextractor} software \citep{1996A&AS..117..393B} directly 
using the MIRI/F1000W map as detection image. We follow the hot-mode extraction described in \citet{2013ApJS..206...10G}, 
which is particularly effective for detecting extremely faint sources and construct a photometric MIRI/F1000W catalog based on the \textsc{sextractor} magnitude estimates \texttt{MAG\_AUTO}.
We cross-match our catalog (using a 0\farcs2 search radius) to the 
official JADES DR2 catalog published by \citet[][]{2023arXiv231012340E} to only consider 
those sources showing a red F444W - F1000W color ($\rm F444W - F1000W \geq 1.5$) and a robust F1000W detection (S/N$~ \geq 5$). 
The reddest object we find in our sample displays a $\rm F444W - F1000W = 2.33 \pm 0.06$. 
We name this MERO {\it Virgil}, to acknowledge the fact that it is very close in projection to {\it Cerberus} (separation $\delta \approx $~2\farcs3). 
An RGB image of the source built with NIRCam and MIRI data is presented in the top right panel of Figure \ref{fig:midis}, together with a series of postage stamps showing {\it Virgil} from 0.7 to 21$\mu$m.

\subsection{Redshift assessment of {\it Virgil}}
\label{sec:redshift}
In the JADES DR2 catalog \cite[][]{Eisenstein+23}, {\it Virgil} is listed with \texttt{ID 206038} and a photometric redshift $z_{phot} \approx 6.62$. 
This redshift estimate agrees with the fact that our source appears to be a NIRCam F090W drop out, see the cutouts presented in Figure~\ref{fig:midis}. 

\begin{figure}
    \centering
    \includegraphics[width = .233\textwidth, trim = {.35cm .35cm, .35cm, .35cm}, clip = True]{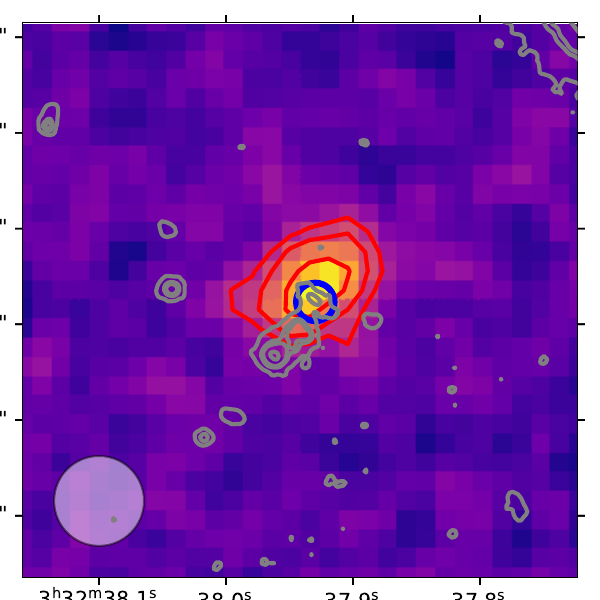}
    \includegraphics[width = .233\textwidth]{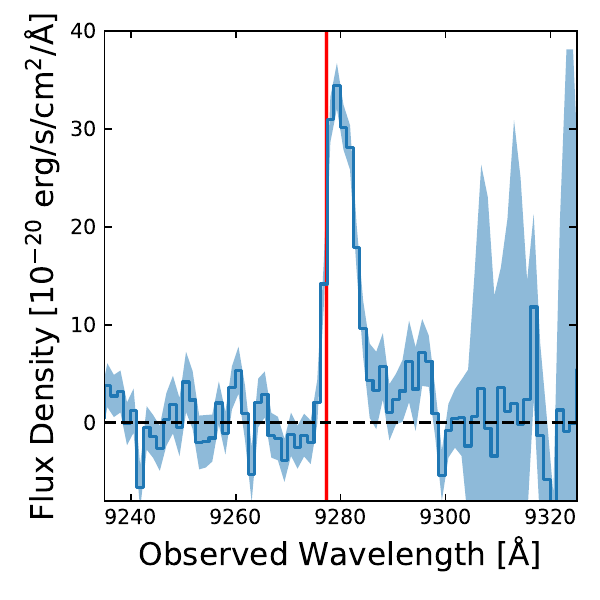}
    \caption{\textbf{Left panel}: MUSE pseudo-narrow-band image ($6'' \times 6''$) of \textit{Virgil}'s Ly$\alpha$ emission. Red contours are drawn at 2, 3 and 5 $\sigma$ while gray contours are representative of the JWST/NIRCam F200W imaging. The source position is indicated with 0.2\arcsec~radius circle. The MUSE PSF is shown in the bottom left corner. \textbf{Right panel}: MUSE 1D spectrum of {\it Virgil}'s Ly$\alpha$. The vertical red line is indicative of the expected position of the Ly$\alpha$ at $z = 6.6312$.
   }
    \label{fig:muse}
\end{figure}

However, to better constrain the properties of our target, we investigate if we can derive a spectroscopic redshift $z_{spec}$ from the available MUSE observations. 
In fact, based on the $z_{phot}$ reported by the JADES DR2 catalog, the MUSE data should cover {\it Virgil}'s rest-frame UV emission. 
Therefore, we look for the possible presence of emission lines, and in particular, of the Ly$\alpha$ line in the MUSE cube. 
To do so, we extract the MUSE 1D spectrum at the coordinates of our target within circular apertures of different sizes ($r = ~$0\farcs4, 0\farcs8). 
Despite being strongly contaminated by the extended rest-frame UV emission of a close-by LAE at $z_{\rm spec} \simeq 4.77$ \cite[\texttt{ID 53};][]{Bacon+23}, 
in the MUSE 1D spectrum we find a spectral feature appearing only at the position of the MIRI extremely red object. 
This line cannot be explained by any known emission line in the UV spectrum of galaxies at $z = 4.77$ and falls at the expected wavelength of the Ly$\alpha$ at $z \simeq 6.63$ (i.e. $\approx 9280$~\AA, see right panel of Figure~\ref{fig:muse}). 
The classification of this spectral feature as {\it Virgil}'s Ly$\alpha$ is in agreement with the publicly available catalog of MUSE-detected sources by \citet{Bacon+23}. 
Particularly, \citet{Bacon+23} list this source (\texttt{ID 7699}) as an LAE at a spectroscopic redshift $z_{spec} = 6.6312\pm 0.0019$ (\texttt{quality\_flag = 2}, 
i.e. good confidence\footnote{An LAE with \texttt{quality\_flag = 2} has a Ly$\alpha$ S/N~$ > 5$ and a width and asymmetry 
compatible with typical Ly$\alpha$ line shapes \citep{Bacon+23};}) with a Ly$\alpha$ flux $F({\rm Ly\alpha}) \approx 1.6 \times 10^{-18}~{\rm erg~s^{-1} cm^{-2}}$ 
(corresponding to a Ly$\alpha$ luminosity $L(\rm Lya) \approx 8.1 \times 10^{41}\rm~ erg ~s^{-1}$) and S/N~$\approx 8$. 
Therefore, we assume $z_{spec} = 6.6312\pm 0.0019$ as {\it Virgil}'s redshift for the rest of our analysis.
Nonetheless, given the wealth of recent spectroscopic campaigns in the HUDF, we also verify if our target had been observed by the FRESCO (PID: 1895; PI: P. Oesch, \citealt{Oesch+23}), the NGDEEP (PID: 2079; PI: Finkelstein, \citealt{Bagley+24}) and JADES surveys.
Unfortunately, our target is not detected in FRESCO's NIRCam Wide Field Slitless Spectroscopy (WFSS) observations\footnote{The FRESCO line sensitivity (estimated for compact sources and integrated over the full extent of the line) is $\approx 2 \times 10^{-18}~\rm erg~s^{-1} cm^{-2}$ at 5$\sigma$ \citep{Oesch+23};}, 
nor in the NGDEEP's NIRISS WFSS observations\footnote{The NGDEEP spectroscopy was carried out in the F115W, F150W and F200W filters reaching  5$\sigma$ integrated emission line limits of 1.2, 1.3 and 1.5$\times 10^{-18}\rm ~erg~s^{-1}~cm^2$, respectively \citep{Bagley+24};}. 
Finally, the NIRSpec Micro Shutter Array (MSA) observations from JADES \citep{DEugenio+24} in HUDF did not cover the area of our target.

\subsection{Removal of {\it Virgil}'s contaminant}
\label{sec:decontamination}
\begin{figure*}
    \centering
    \includegraphics[width = \textwidth]
    {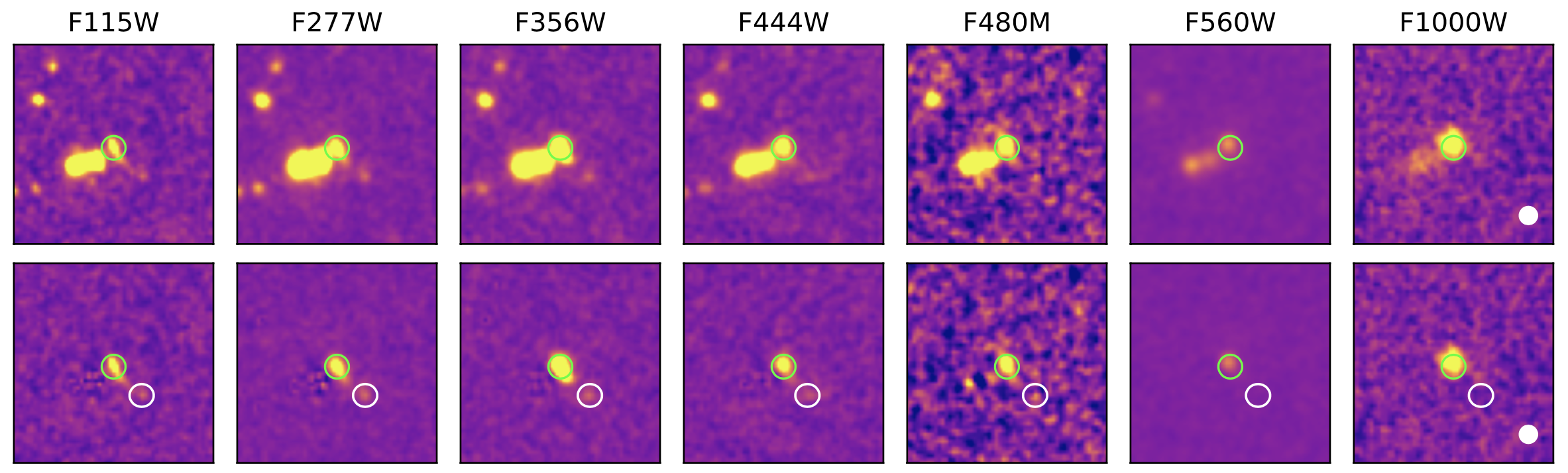}
    \caption{\textbf{Top panels}: cutouts ($3\farcs4\times 3\farcs4$) of {\it Virgil} from NIRCam (F115W, F277W, F356W, F444W, F480M) and MIRI (F560W, F1000W) imaging. We highlight the position of {\it Virgil} with a green circle (0\farcs2 radius). The south-eastern component that gets dimmer at longer wavelengths is the LAE contaminant at $z = 4.77$ \citep{Bacon+23, Matthee+22}. The white circle at the bottom right corner of the F1000W cutouts is representative of the MIRI 10$\mu$m PSF (FWHM~$\simeq 0\farcs33$). 
    \textbf{Bottom panels}: same cutouts after modeling and subtracting the contaminant at $z = 4.77$ and close-by sources. The white circle (0\farcs2 radius) highlights the position of {\it Beatrix}.} 
    \label{fig:degraded_ima}
\end{figure*}

In the JADES DR2 catalog, {\it Virgil}'s contaminant at $z \approx 4.77$ (\citealt{Bacon+23}; see also \citealt{Matthee+22}) turns out to be constituted by two sources: \texttt{ID 285736} 
and \texttt{ID 206035}. 
With respect to {\it Virgil}'s location, \texttt{ID 285736} is about 0\farcs4 away while \texttt{ID 206035} has a separation of 0\farcs7 both in the south-east direction.
To avoid the contamination of {\it Virgil}'s photometry due to this extended lower redshift LAE, we model and subtract its contribution at all wavelengths.
To do so, we resort to the galaxy morphology modeling tool \textsc{AstroPhot} \citep{Stone+23}.
With \textsc{AstroPhot}, we model \texttt{ID 285736} and \texttt{ID 206035} with single S\'ersic profiles while masking {\it Virgil}.  
From the first results, we find that to improve the quality of the residuals in several filters, it is necessary to introduce an optional third S\'ersic component between \texttt{ID 285736} and \texttt{ID 206035}, see also \cite{Matthee+22}.
In Figure~\ref{fig:degraded_ima}, we present the residuals of the best-fit models obtained with \textsc{AstroPhot} for some of the available NIRCam and MIRI filters.

\subsection{{\it Virgil}'s morphology}
\label{sec:morphology}
After the removal of \textit{Virgil}'s contaminant and before extracting its photometry (see \S\ref{sec:decontamination}), we briefly investigate {\it Virgil}'s morphology. 

At NIRCam wavelengths our target is clearly elongated in the south-west direction and features a tail that seems to end up in a (faint) clump ({\it Beatrix}; $\alpha$(J2000.0) = 03:32:37.8844 (hours); $\delta$(J2000.0) = -27:47:10.979) that can be detected at several NIRCam wavelengths but is completely missing in the MIRI imaging, see Figure~\ref{fig:degraded_ima}. 
On the basis of the current data, it is difficult to assess whether {\it Beatrix} is a clump of {\it Virgil} or a different object at a different redshift (see \S\ref{sec:photometry}).
However, if confirmed at {Virgil}'s redshift, the separation between {\it Virgil} and {\it Beatrix} ($\delta \approx ~$0\farcs75) would correspond to a projected distance of about 4 pkpc and, considering its point-like nature, a size $r \lesssim 0.''06 \simeq 0.3$~pkpc. 

\begin{figure*}
    \centering
    \includegraphics[width = \textwidth]{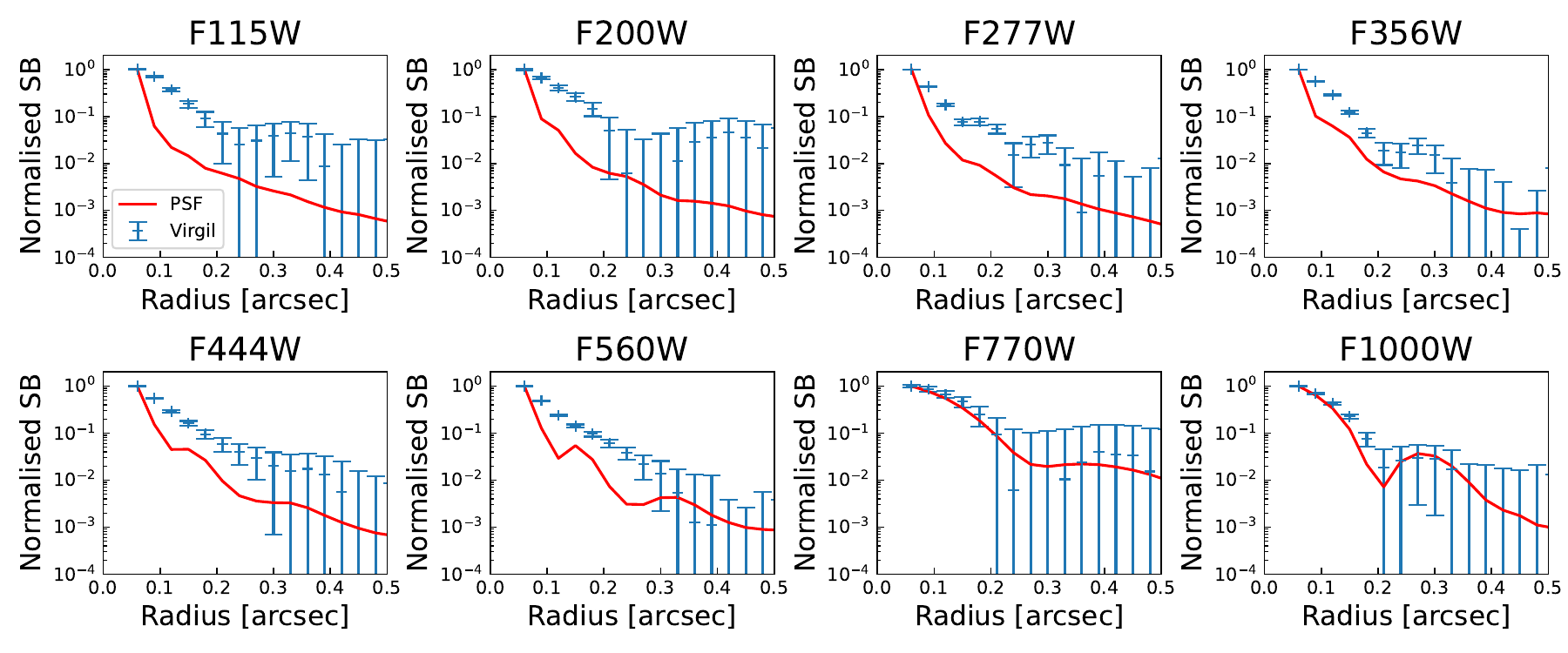}
    \caption{Surface brightness (SB) profiles of {\it Virgil} at different NIRCam and MIRI wavelengths (blue errorbars). The surface brightness profiles are normalised at {\it Virgil}'s peak. The red curves show the JWST PSF trend for the different instruments and filters.}
    \label{fig:sb-profile}
\end{figure*}

After masking {\it Beatrix} and other close-by sources, the extracted surface brightness profile shows that {\it Virgil} is an extended object at wavelengths $\lambda < 7~\mu$m, while at 7.7 and 10~$\mu$m, it is unresolved due to the coarser MIRI PSF . 
This is confirmed by the extraction of {\it Virgil}'s radial surface brightness profile and its comparison with the PSF trend in the different NIRCam and MIRI filters, see Figure~\ref{fig:sb-profile}. 

If we fit a S\'ersic profile \citep{Sersic+63} to the SB radial profiles of the different broad-band NIRCam and MIRI filters covering {\it Virgil}'s rest-frame optical emission (i.e. F200W, F277W, F444W, F560W), we find a median value of the S\'ersic index $n = 0.93^{+0.85}_{-0.31}$ (16th and 84th percentiles) and an effective radius $r_e \simeq 0.''09^{+0.01}_{-0.02} = 0.49^{+0.05}_{-0.11}\rm~pkpc$ (16th and 84th percentiles).
Both parameters are in agreement with the expected values of galaxy sizes and S\'ersic indices at $z \simeq 6$ \citep{Sun+24}.

\subsection{Extraction of {\it Virgil}'s photometry}
\label{sec:photometry}

After having removed the lower redshift interloper from all the 31 available HST (ACS and WFC3) and JWST (NIRCam and MIRI) bands, we extract {\it Virgil}'s photometry in the wavelength range 0.4\,--\,25~$\mu$m with \textsc{sep} \citep{Barbary+17}, a \textsc{Python} version of \textsc{SExtractor} \citep{Bertin+96}. 
However, at the longest MIRI wavelengths ($> 15~\mu$m), the available imaging is too shallow to detect {\it Virgil}.
Also after stacking together all the different MIRI bands available above 15 $\mu$m (i.e. F1800W, F2100W, F2550W) we fail to detect our target. 
Hence, above 15 $\mu$m we make use of the 3$\sigma$ upper limits on the flux density. 
 
We extract Virgil photometry $\leq 15~\mu$m within \texttt{Kron apertures} \citep{Kron+80}, allowing the Kron parameters to vary from filter to filter and replicating standard \textsc{SExtractor} settings, i.e. $K = 2.5$ and a minimum radius $r = 1\farcs75$.
Then, we correct the extracted fluxes to account for the missing flux outside the Kron aperture.
To do so, we compute the fraction of the missing light outside the Kron apertures \cite[e.g.][]{Whitaker+11, Weaver+23a,  Kokorev+24} in comparison to the curve of growth of the different PSFs.
For the JWST filters, we employ the PSF models by {\sc WebbPSF} \citep{Perrin+14}, while for HST we resort to the curve of encircled energy reported by STScI\footnote{The encircled energy curves are available at \url{https://www.stsci.edu/hst/instrumentation/};}.
We finally correct for Galactic extinction. 
Following \cite{Iani+24}, we obtain the Galactic extinction corrections assuming the color excess reported on the IRSA webpage\footnote{\url{https://irsa.ipac.caltech.edu/applications/ DUST/}; 
} at {\it Virgil}'s coordinates \cite[$\rm E(B - V) = 0.008$, ][]{Schlafly+11} and, for all filters with an effective wavelength $\lambda_{\rm eff} < 1.25~\rm \mu m$, applying the \cite{fitzpatrick+99} extinction law, while at longer wavelengths ($1.25 \leq \lambda_{\rm eff} < 8~\rm \mu m$), resorting to the \cite{indebetouw+05} law. 
The correction factors for Galactic extinction are, however, quite negligible ($< 3\%$).

As in the case of \textsc{SExtractor} \citep{Sonnett+13}, \textsc{sep} tends to underestimate the errors on the fluxes.
To derive more reliable errors, we estimate them drawing in each filter 1000 random Kron apertures on the sky region around {\it Virgil} ($6''\times 6''$) and, after having masked all the close-by sources, measuring their fluxes. We then take the standard deviation of all the different measurements taking into account effects of pixel correlation \citep{Fruchter+02}.
Similarly to the fluxes, we also correct the errors to total flux and Galactic extinction.
Finally, we impose as a minimum error on the photometry a value of 0.05 mag for all bands.
We present {\it Virgil}'s photometry in Table~\ref{tab:photo_virgil}.

\begin{figure}
    \centering
    \includegraphics[width = .23\textwidth]{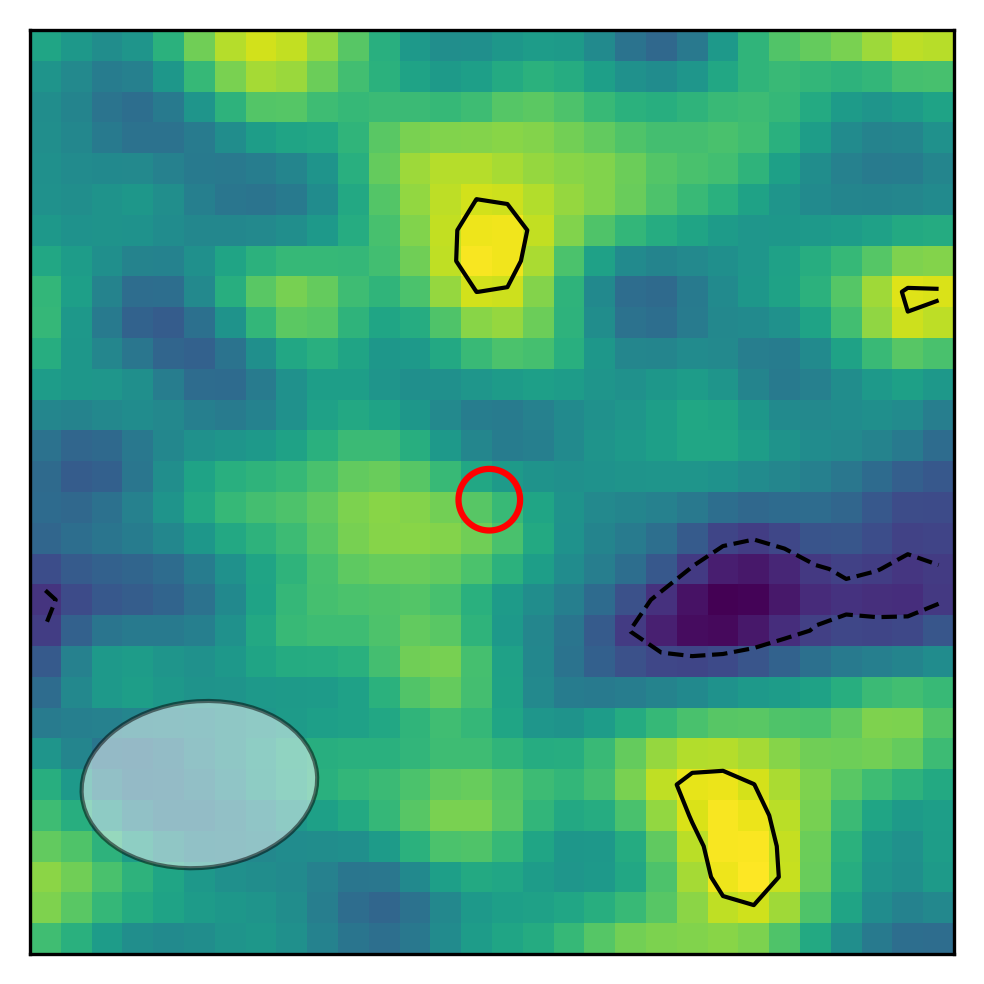}
    \includegraphics[width = .23\textwidth]{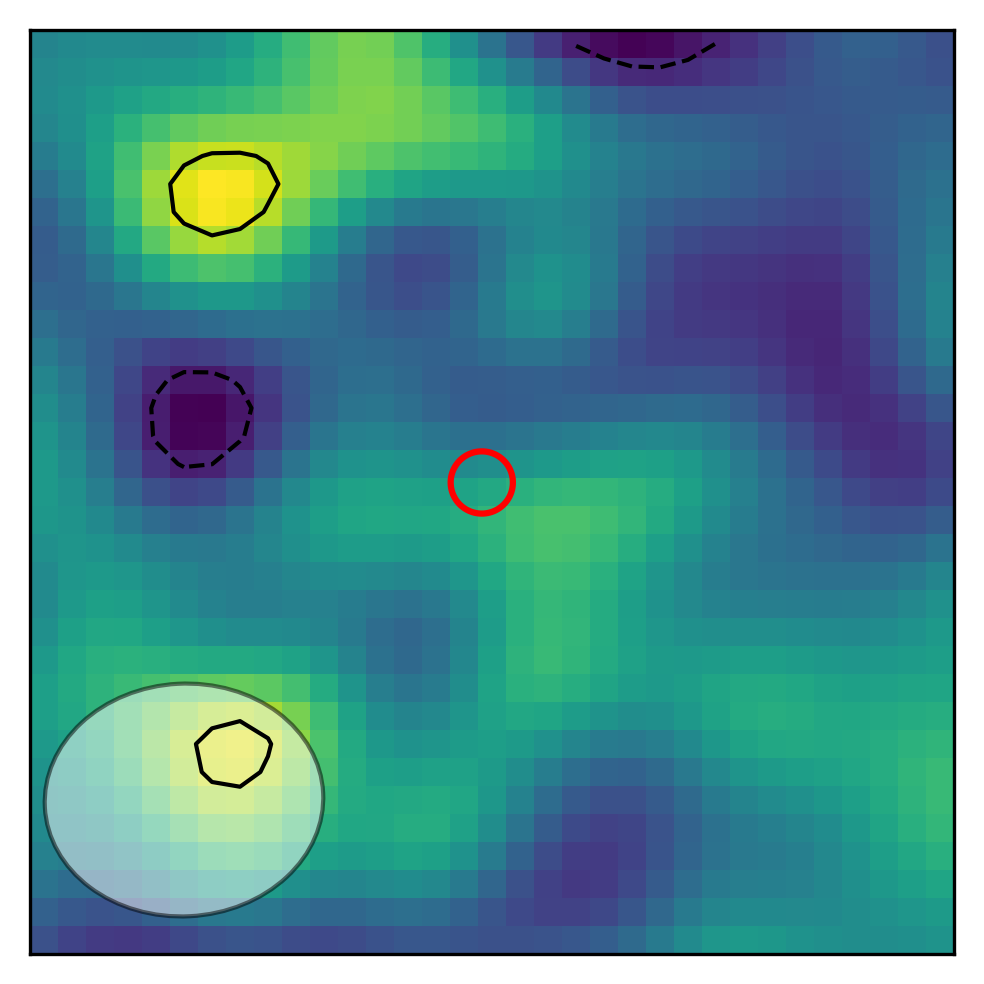}
    \caption{ALMA cutouts ($6'' \times 6''$) at 1.2 mm (band 6) and 3.0 mm (band 3) from ASPECS. Contours are drawn at 1$\sigma$ intervals starting at $\pm 2 \sigma$ (dashed lines show negative contours). The source position is indicated with a 0.2\arcsec~radius circle. The beam size is shown in the bottom left corner. The source is not detected at either wavelength.}
    \label{fig:alma}
\end{figure}

In addition to HST and JWST imaging, we search for detection of {\it Virgil} at (sub-)millimeter wavelengths leveraging on the available deep ASPECS ALMA dust continuum imaging at 1.2 and 3.0~mm. 
By carefully inspecting these maps, {\it Virgil} is not detected at either wavelengths, see Figure~\ref{fig:alma}. 
This finding implies 3$\sigma$ upper limits on the flux density at 1.2 mm and 3.0 mm of 27.9 $\mu$Jy and 4.2 $\mu$Jy, 
respectively \cite[assuming it is an unresolved point-like source;][]{Boogaard+23, Perez-Gonzalez+24b}.

To assess the quality of our photometry on the images cleaned for the contaminant, we decide to test if, by running Spectral Energy Distribution (SED) fitting codes on our photometry, we retrieve photometric redshifts in agreement with the spectroscopic redshift from MUSE. 
To this purpose, we employ both \textsc{Eazy} \citep{Brammer+08} and \textsc{CIGALE} \citep{Burgarella+05, Noll+09, Boquien+19}. 
For {\sc Eazy}, we adopt the v1.3 templates, which include a dusty galaxy with a high-equivalent width emission line spectrum, as well as new models for LRDs and high-redshift AGN+torus recently added based on JWST data \citep{Killi+23}. 
We do not impose any prior and work with minimum $\chi^{2}$ photometric redshift estimates in the redshift range $z = 0-20$ and consider {\it Virgil}'s photometry up to 10~$\mu$m.
For \textsc{CIGALE}, we run it only considering stellar templates and HST + JWST/NIRCam photometry, thus excluding the red upturn of our target's photometry. 
We use the same setting as presented in \S~\ref{sec:sed}.
According to both codes, the probability distribution of the photometric redshifts (PDZ) is uniquely determined (one single narrow peak) to be at $6.61\pm 0.02$ (\textsc{Eazy}) and $6.68 \pm 0.23$ (\textsc{CIGALE}). 
Overall, we find a very good agreement with the spectroscopic estimate from MUSE. 

In addition to {\it Virgil}, we extract the photometry for {\it Beatrix} following the same methodology as presented above.
Our aim is to understand if we can assess the redshift of this object and discern if it is part of {\it Virgil} (clump) or a source at a different redshift.
Due to the faintness of the source, however, we can only detect {\it Beatrix} in few HST and JWST/NIRCam bands.
On the derived photometry, we run both \textsc{Eazy} and \textsc{Cigale}. 
Both codes return a multi-peaked PDZ. Interestingly, one of the peaks falls at {\it Virgil}'s redshift and another one corresponds to the redshift of the contaminant LAE at $z \simeq 4.77$. 
The other peaks, however, suggest even lower redshift solutions ($z_{phot} < 2$).
Due to the limited dataset available and the fact that we are not sure about its redshift, we do not further investigate this object.

\begin{deluxetable}{lccc}[t]
    \tablenum{1}
    \tablecaption{Photometry of \textit{Virgil}}
    \tablewidth{0pt}
    \tablehead{ 
        \colhead{Instrument} & \colhead{Filter} & \colhead{$f_{\nu}$} & \colhead{err($f_{\nu}$)} \\
         &   & \colhead{[nJy]} & \colhead{[nJy]} 
    }
\startdata
JWST/NIRCam & F090W  & 12.86  &   2.25  \\                   
HST/ACS\_WFC & F850LP &  3.38  &   2.70  \\                  
HST/WFC3\_IR & F105W  & 22.07  &   1.95  \\                  
JWST/NIRCam & F115W  & 30.54  &   2.00  \\                  
HST/WFC3\_IR & F125W  & 27.06  &   2.51  \\                  
HST/WFC3\_IR & F140W  & 22.90  &   2.59  \\                  
JWST/NIRCam & F150W  & 28.92  &   1.85  \\                             
HST/WFC3\_IR & F160W  & 28.52  &   2.82  \\                  
JWST/NIRCam & F182M  & 24.63  &   2.65  \\                  
JWST/NIRCam & F200W  & 30.32  &   2.04  \\                  
JWST/NIRCam & F210M  & 36.70  &   3.10  \\                  
JWST/NIRCam & F277W  & 40.11  &   1.85  \\                   
JWST/NIRCam & F335M  & 38.88  &   2.26  \\                             
JWST/NIRCam & F356W  & 93.84  &   4.32  \\                   
JWST/NIRCam & F410M  & 37.01  &   1.89  \\                   
JWST/NIRCam & F430M  & 37.79  &   4.80  \\                              
JWST/NIRCam & F444W  & 49.33  &   2.27  \\                   
JWST/NIRCam & F460M  & 46.99  &   6.25  \\                   
JWST/NIRCam & F480M  & 67.58  &   7.91  \\                   
JWST/MIRI   & F560W  & 107.70 &   3.03  \\                    
JWST/MIRI   & F770W  & 248.20 &  30.13  \\                   
JWST/MIRI   & F1000W & 399.10 &  23.79  \\                   
JWST/MIRI   & F1280W & 382.70 & 161.10  \\                   
JWST/MIRI   & F1500W & 379.50 & 296.20  \\   
\enddata
    \tablecomments{In the above table, we report the photometry of {\it Virgil} as extracted after the modeling of the $z = 4.77$ LAE contaminant (see \S\ref{sec:decontamination}). The table is limited to only those bands where it was possible to detect {\it Virgil}.}
    \label{tab:photo_virgil}
\end{deluxetable}

\subsection{Possible galaxy - galaxy lensing effect on {\it Virgil}}
Due to the closeness of the contaminant to {\it Virgil} and the extended and stretched morphology of our target, we investigate if {\it Virgil} could be affected by a galaxy - galaxy lensing effect \cite[e.g.][]{Matthee+17}. 
To estimate the magnification of {\it Virgil} by the foreground LAE we resort to the usage of the \textsc{lenstronomy} software \citep{Birrer+18}.  
The measured stellar masses of the two components of the foreground LAE are $M_\star \simeq 10^{8.3}\rm~M_\odot$ and $10^{8.6}\rm~M_\odot$ \citep{Eisenstein+23} for the objects located 0\farcs4 and 0\farcs7 from {\it Virgil}, respectively. 
The two components are located 0\farcs32, or about 2.1 pkpc from each other at $z = 4.77$.
The separations involved are sufficiently large that the lensing model is mostly unaffected by assumptions on the mass profile of the lensing galaxies; the Einstein radius is $\sim 5$~mas using a point-mass approximation, and $\lesssim 20$~mas when assuming extended mass distributions (maximized using an ellipsoid configuration). 
This configuration results in a magnification $< 5$\% of {\it Virgil} even under extreme assumptions for the total-mass-to-stellar-mass ratios of 10 for both foreground galaxies.
Due to the minimum value of the retrieved magnification, we do not consider corrections for lensing in the following analysis.

\subsection{Spectral energy distribution analysis of {\it Virgil}}
\label{sec:sed}

After having extracted its decontaminated photometry, we try to reproduce {\it Virgil}'s observed SED with different SED fitting codes, namely, {\sc CIGALE}
,  {\sc Bagpipes} \citep{Carnall+18}, {\sc prospector} \citep{Johnson+21}, and {\sc synthesizer-AGN} \citep{Perez-Gonzalez+03, Perez-Gonzalez+08, Perez-Gonzalez+24a}. 
Due to {\it Virgil}'s red F444W - F1000W color, we decide to run the codes both excluding and considering the presence of an AGN (see \S~\ref{sec:discussion}).
For the SED fitting, we consider all the photometry available, including the ALMA upper limits at 1.2 and 3mm. 

We first employ {\sc CIGALE} assuming a delayed exponentially declining SFH (delayed-$\tau$ model) modelled with two stellar populations. 
We adopt the \citetalias{Bruzual+03} stellar populations models with both solar and sub-solar metallicity ($Z = 0.2 Z_\odot$), and the Chabrier IMF. 
We include nebular continuum and emission lines using solar and sub-solar metallicity and allowing the electron density $n_e$ and ionisation 
parameter $U$ to assume values of $n_e = 10, 100, 1000\rm~cm^{-3}$ and $\log_{10}(U) = -2,-1$, respectively. 
For the dust attenuation, we adopt \citetalias{Calzetti+00} while for the far-IR emission, we resort to the \cite{Draine+14} models. 
We add the AGN emission using the SKIRTOR models \citep{Stalevski+12, Stalevski+16} following the initial parameters suggested by \cite{Yang+23} but allowing for the presence of both a Type I (unobscured) and 2 (obscured) AGN.

As a second SED-fitting code, we run {\sc Bagpipes}, a stellar population synthesis modeling package built on the \citetalias{Bruzual+03} 
spectral library with the 2016 version of the MILES library \citep{Falcon-Barroso+11}. 
The code uses a \cite{Kroupa+01} IMF, adopts the \citetalias{Calzetti+00} dust attenuation curve and includes nebular emission lines. 
The ionisation parameter $\log_{10}(U)$ is set to vary between $(-4,\, -2)$. 
The SFH is set to two components delayed-$\tau$ model and we included an AGN component as in \cite{Carnall+23}.

For {\sc prospector} we adopt the setup described in detail in \cite{Langeroodi+23} and \cite{Langeroodi+23b}. In brief, we use 5 temporal bins to model the star-formation history non-parametrically while applying the continuity prior from \cite{Leja+19}. The nebular emission is modelled using the \textsc{cloudy} \citep{Chatzikos+23} templates compiled in \cite{Byler+17}, while the gas-phase metallicity, stellar metallicity, and ionization parameters are modelled as free parameters. We model the dust attenuation using the two-component model of \cite{Kriek+13}, where one component affects the entire galaxy and the other models the additional reddening at the birthplace of young stars. 

Finally, we run {\sc synthesizer-AGN}. The code assumes that the SED can be modelled with a composite stellar population 
\citep{Perez-Gonzalez+03, Perez-Gonzalez+08} and AGN emission coming from the accretion disk and the dusty torus \citep{Perez-Gonzalez+24a}. 
The stellar emission includes a young and a more evolved star formation event, each one described by a delayed exponential 
function with timescales between 1 Myr and 1 Gyr, and with ages from 1 Myr up to the age of the Universe at the redshift of the source.
The attenuation of the emission from each stellar population is independent and described by the \citetalias{Calzetti+00} law, 
with A(V) values ranging from 0 to 10 mag for each population, considered to have completely independent attenuation. 
The stellar emission is described by the \citetalias{Bruzual+03} models, assuming a Chabrier IMF with stellar mass limits 
between 0.1 and 100 M$_\odot$, and the nebular emission is also considered \citep{Perez-Gonzalez+03}. 
The AGN emission is modelled with a QSO average spectrum \citep{Vanden-Berk+01, Glikman+06}. 
The dust emission from the AGN is modelled with the self-consistent templates of AGN tori presented in \cite{Siebenmorgen+15}.

We present and discuss the results derived from the different SED-fitting codes in  \S\ref{sec:discussion}.

\begin{figure*}
    \centering
    \includegraphics[width = .8\textwidth]{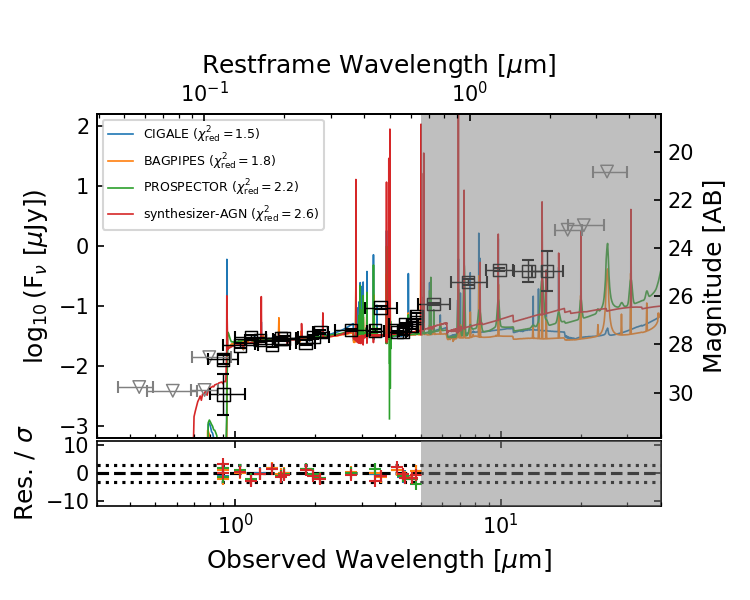}
    \caption{{\it Virgil}'s photometry (black open squares and gray triangles for the upper limits) and best-fit models obtained with the SED fitting codes {\sc CIGALE} (blue) and {\sc Bagpipes} (orange) 
    when limiting the analysis to the HST + JWST/NIRCam coverage, i.e. up to about 5~$\mu$m (rest-frame 0.65~$\mu$m). The gray shaded area is indicative of rest-frame wavelengths above 0.65~$\mu$m. The SED fittings performed in this case do not include any AGN contribution. The bottom panel shows the ratio between the residuals (i.e. the synthetic minus the observed photometry) and the photometric error $\sigma$ for all the available bands. The dotted lines are indicative of residuals equal to $\pm 3\sigma$.}
    \label{fig:sed_blue}
\end{figure*}

\begin{figure*}
    \centering
    \includegraphics[width = .77\textwidth]{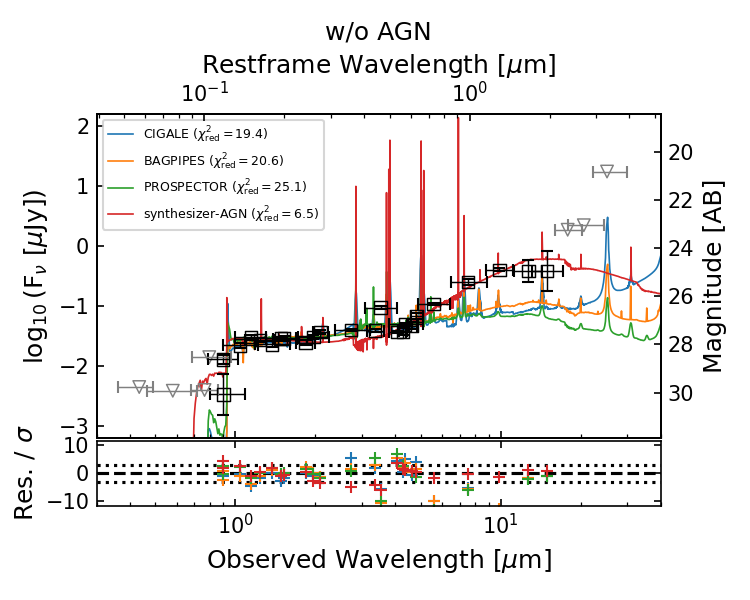}\\ 
    \includegraphics[width = .77\textwidth]{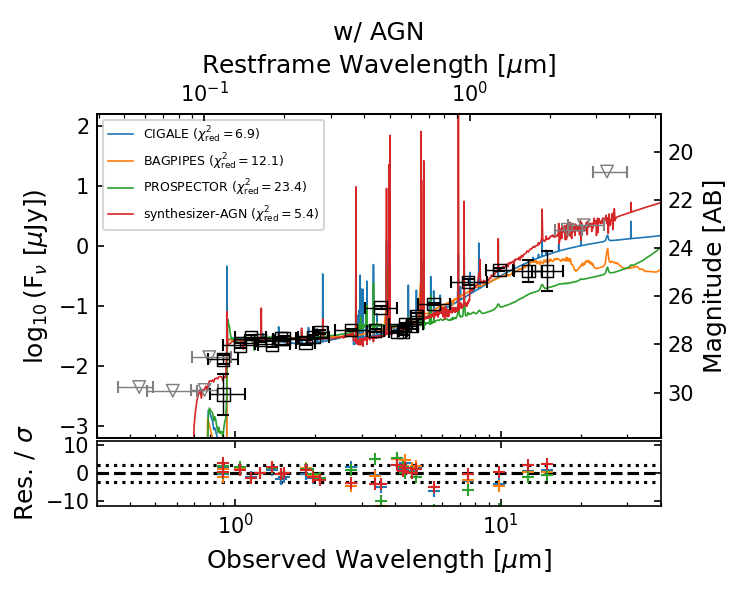} 
    \caption{{\it Virgil}'s photometry (black open squares and gray triangles for the upper limits) and best-fit models obtained with the SED fitting codes {\sc CIGALE} (blue), {\sc Bagpipes} (orange), {\sc prospector} (green) and  {\sc synthesyzer-AGN} (red) in the case of excluding (upper panel) and assuming (lower panel) an AGN component. The bottom panel shows the ratio between the residuals and the photometric error $\sigma$ for all the available bands. The dotted lines are indicative of residuals equal to $\pm 3\sigma$. For the sake of clarity, we limit the above panels to wavelengths $< 40~\mu$m even if the ALMA upper limits were taken into account during the fitting procedure.}
    \label{fig:virgil_sed}
\end{figure*}

\section{Discussion}
\label{sec:discussion}
\subsection{{\it Virgil}'s properties without MIRI}
\label{sec:short_sed}
As a first step, we decide to investigate {\it Virgil}'s properties based only on its HST and JWST/NIRCam photometry, i.e. limiting our study to wavelengths $< 5~\mu$m.

The SED of {\it Virgil} reveals a rest-frame blue UV continuum. 
Following \cite{Castellano+12}, by fitting a power law to the observed fluxes ($f_\lambda \propto \lambda^{\beta_{\rm UV}+2}$), we estimate a UV continuum slope of $\beta_{\rm UV} = -2.1 \pm 0.3$ and an absolute UV magnitude at 1500\AA~M(UV)~$ =-19.1\pm 0.1$~mag.
Both parameters are well in agreement with what was recently found for other LAEs at similar redshifts \citep{Iani+24} and suggest negligible effects due to dust extinction \cite[$A_{\rm 1500} < 0.3$~mag; e.g.,][]{Meurer+99}. 
If we convert the M(UV) into SFR following the prescription by \cite{Kennicutt+12}, we find an SFR(UV)$~\simeq 1.7~\rm M_\odot~yr^{-1}$ (not corrected for dust extinction).

At optical wavelengths, {\it Virgil}'s SED is characterised by a clear photometric excess in the F356W band (F335M $-$ F356W~$=0.96 \pm 0.08$).
Due to {\it Virgil}'s redshift (i.e., $z_{spec} = 6.6312 \pm 0.0019$), this excess can be explained as the direct consequence of a strong  H$\beta$ + [\ion{O}{3}]$\lambda\lambda4959,5007$ ([\ion{O}{3}], hereafter) emission line complex entering the F356W filter. 
If we assume the continuum below H$\beta$ + [\ion{O}{3}] as the average value between the F356W adjoining filters F335M and F410M and apply the prescriptions by \cite{Marmol-Queralto+16}, we can convert the photometric excess in F356W into the rest-frame equivalent width EW$_0$ of the line complex.
From our photometry, we estimate EW$_0({\rm H}\beta + [$\ion{O}{3}$]) \approx 560$~\AA.
This value is in good agreement with recent literature for emitters at $z \gtrsim 6$ \cite[][]{Endsley+21, Prieto-Lyon+23, Rinaldi+23a, Matthee+23, Caputi+24}.
We highlight that an increase in flux is also observed in the NIRCam longest wavelength filter, e.g. F460M $-$ F480M~$= 0.39 \pm 0.19$.
At {\it Virgil}'s redshift and without the information coming from the MIRI imaging, this flux excess could be easily ascribed to the H$\alpha$ emission line entering the F480M filter. 
In fact, medium band filters are known for their heightened sensitivity to strong emission lines \cite[e.g.][]{Papovich+23, Rinaldi+23a, Caputi+24}.

Hence, by limiting the analysis of {\it Virgil} at rest-frame wavelengths $\leq 0.65~\mu$m, the observed SED of {\it Virgil} is well represented by a typical LAEs' SED with strong optical emission lines \cite[e.g.,][]{Ouchi+20}.
This is also confirmed by our runs with SED fitting codes only considering the available photometry below the MIRI/F560W imaging. 
In this case, all codes model {\it Virgil} as a young ($\leq 100$ Myr) star-forming ($\rm SFR = 1 - 2~ M_\odot~yr^{-1}$) low-mass galaxy ($M_\star \simeq 10^{8.0-8.5}\rm ~M_\odot$) with low dust content ($\rm A_V \leq 0.8$~mag), see Figure~\ref{fig:sed_blue}. 

\subsection{{\it Virgil}'s full SED}
\label{sec:full_sed}
The distinctive feature of {\it Virgil} lies in its pronounced increase in brightness beyond 0.6~$\mu$m rest-frame (4.8~$\mu$m observed-frame), leading to a notable upturn in its SED at near-IR wavelengths.
By fitting a power law ($f_\lambda \propto \lambda^{\alpha_{\rm NIR}}$) to the observed fluxes in the wavelength range $4 - 10\mu$m (observed), we find a spectral index $\alpha_{\rm NIR} = 2.8 \pm 0.1$.
While the increase in the reddest NIRCam medium band filters F460M and F480M could be potentially caused by the presence of a strong  H$\alpha$ + [\ion{N}{2}]$\lambda\lambda6548,6584$ 
line complex \cite[e.g.,][]{Papovich+23}, the steep rising MIRI SED from 0.7~$\mu$m (F560W) up to 1.3~$\mu$m (F1000W) rest-frame cannot be solely explained by strong emission lines. 
In fact, although at $z = 6.6312$ the Paschen-$\beta$ (Pa$\beta$) emission line enters the F1000W band, atomic physics predicts that the Pa$\beta$ intensity is about 1/20 of the H$\alpha$ line \cite[assuming case B recombination, e.g.][]{Osterbrock+06}. Besides, F1000W is a very wide band, with an effective width of about 17000~\AA. 
Hence, for a galaxy at {\it Virgil}'s redshift, already an excess of 0.1 mag in F1000W would imply a rest-frame equivalent width for the emission line of about 200~\AA, i.e. roughly 1/3 of what we estimate for the H$\beta$ + [\ion{O}{3}] complex. 
Based on our previous best SED fitting models 
(see \S~\ref{sec:short_sed} and Figure~\ref{fig:sed_blue}), we would expect an excess of more than 2.0 mag with respect to the models' continuum in F1000W, 
thus implying an unrealistically strong Pa$\beta$ line.  
A similar reasoning can be applied to {\it Virgil}'s detection (S/N~$\approx 5$) in F770W (effective width of about 18300~\AA), with a flux $\approx$~2.5 times higher than what we detect in F560W (F560W - F770W~$= 1.01 \pm 0.23$). 

Having excluded that the red upturn is driven by strong emission lines, we fit {\it Virgil}'s observed photometry (including the ALMA upper limits at 1.2 and 3mm) in a two-fold way: assuming that its SED is uniquely due to the contribution of stellar populations and the ISM,  and adding the effects of an AGN.
In fact, {\it Virgil}'s SED steep rise at the rest-frame near-IR domain of galaxy spectra could be the sign of an AGN dusty torus \cite[e.g.][]{Hickox+18}.

\subsubsection{Only Stellar Populations models}
\label{subsubsec:only_stars_fit}
\begin{figure}
    \centering
    \includegraphics[width = .5\textwidth]{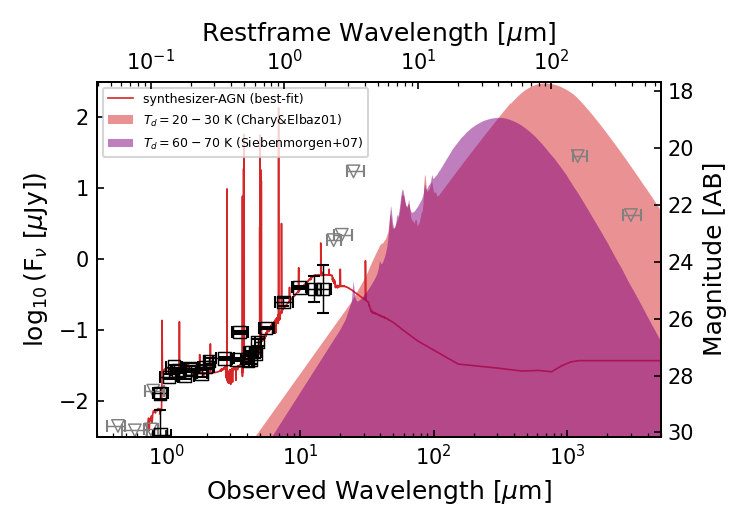}
    \caption{Decomposition of the dust component from {\it Virgil}'s photometry (black open squares and gray triangles for the upper limits) for the best-fit model obtained with the SED fitting code {\sc synthesizer-AGN} (red). The red-shaded area is indicative of the model of dust emission with temperatures of 20 - 30 K \citep{Chary+01} for a $L_{\rm dust} \approx 10^{11.5}\rm~L_\odot$ (as predicted from {\sc synthesizer-AGN}). The purple-shaded area represents a model with higher dust temperature \cite[60 - 70 K, ][]{Siebenmorgen+07}. In the case of two populations and strong attenuation, only models with higher $T_{\rm dust}$ can comply with the ALMA upper limits.}
    \label{fig:sed_dust_component}
\end{figure}

When implementing only stellar populations (see Table~\ref{tab:sed-no-agn}), in general, we struggle to reproduce {\it Virgil}'s overall SED and behaviour at the MIRI wavelengths.  
The only code that can reproduce {\it Virgil}'s photometry and its red MIRI slope with solely stellar populations is {\sc synthesizer-AGN} ($\chi^2_{\it red} = 6.9$) thanks to a combination of two sub-solar ($Z \simeq 0.4 Z_\odot$) young populations (6, 26~Myr old) with very different extinctions ($A_V = 0.5, 4.0$~mag) and combined with a very strong contribution from the nebular continuum, see Figure~\ref{fig:virgil_sed} (top panel).
Due to the presence of the heavily attenuated second stellar component, the total mass of {\it Virgil} retrieved by {\sc synthesizer-AGN} is $\rm M_\star = 10^{9.7\pm 0.1}~M_\odot$.
From the best-fit model, we also derive a total $\rm SFR(UV) \approx 69~M_\odot~ yr^{-1}$ (corrected for dust extinction). 
This solution would suggest a scenario for {\it Virgil} as a dusty starburst galaxy. 
In this case, the blue and red components of the SED would arise from different parts of the galaxy.
The relatively small angular size of {\it Virgil} together with the coarser MIRI resolution and the presence of the lower redshift contaminant (\S\ref{sec:decontamination}) prevent us, however, from further investigating this scenario based on the currently available data.
Nonetheless, we highlight that in this scenario, classical templates of dust emission \cite[e.g., ][]{Chary+01} with dust temperatures $T_{\rm dust} = 20 - 30$~K dominating the dust mass would not comply with the ALMA upper limits for a total absorbed energy of $L_{\rm dust} \approx 10^{11.5}\rm~L_\odot$.
Only assuming models with higher dust temperatures \cite[$\approx 60 - 70$~K; e.g., ][]{Siebenmorgen+07, Schreiber+18, Sommovigo+20, Sommovigo+22} would allow to meet the upper limits at wavelengths $> 1$~mm, see Figure~\ref{fig:sed_dust_component}.

The flexibility of {\sc synthesizer-AGN} in finding such best-fit result is hardly retrievable with the other SED-fitting codes at our disposal and which struggle to return good best-fit models ($\chi^2_{\it red} \gtrsim 20$), failing in reproducing the photometry at the MIRI wavelengths.
In this context, {\sc CIGALE},  {\sc Bagpipes} and {\sc prospector} find very similar best-fit parameters, i.e. $M_\star \simeq  10^9\rm ~M_\odot$, $\rm SFR \simeq 3~ M_\sun~ yr^{-1}$, $A_V \leq 0.8\rm~ mag$ and a mass-weighted age of about 300 Myr.

Interestingly, if we try to fit a graybody emission ($f_\lambda \propto \lambda^{-\beta -3} /(e^{\frac{hc}{\lambda k_{\rm B} T}}-1)$, e.g. \citealt{Caputi+13}) to the residuals of the MIRI datapoints, i.e. after subtracting the best-fit models derived by these three different codes to the observed photometry, we find a graybody temperature $T = 1762^{+177}_{-143}\rm~ K$ (if we assume $\beta = 1.5$). 
Such high value of the graybody temperature is comparable to the typical values for the sublimation of dust \cite[$T_{\rm sub} \approx 1500 - 2000\rm K$ depending on the dust composition, e.g. ][]{Temple+21} and is hardly reconciliable with heating due to normal stellar activity \cite[even in the case of very compact, intense starbursting regions; e.g., ][]{DeRossi+18, Sommovigo+20}.

\subsubsection{Stellar Populations and AGN models}
\label{subsubsec:agn_fit}
To improve the best-fit models, we consider adding an AGN component to the fits. 
As a first step, we verify if {\it Virgil} is an already known AGN in the literature. 
To this purpose, we look for possible counterparts of {\it Virgil} in the available AGN catalogs in HUDF based on both observations by Chandra \cite[e.g.][]{Luo+17, Evans+20} 
and XMM-Newton \citep{Ranalli+13} but find no X-ray counterpart. 

To assess if {\it Virgil} is a weak X-ray emitter and its absence in the X-ray catalogs is not merely due to the depth of the available Chandra and XMM-Newton observations, we estimate an upper limit over its rest-frame X-ray to optical-UV ratio $\alpha_{\rm ox} = -\log_{10}[L({\rm 2keV}) / L({\rm 2500 \AA})]/2.605$ \cite[e.g., ][]{Tananbaum+79, Lusso+10}. 
While we derive {\it Virgil's} luminosity at 2500\AA~ $L({\rm 2500 \AA})$ directly from the available JWST/NIRCam photometry ($L({\rm 2500 \AA}) \approx 1.98 \times 10^{28}\rm~erg/s/Hz$), we infer the luminosity at 2keV $L({\rm 2keV})$ by considering three times the limiting flux of the deepest X-ray observations (i.e. Chandra) in the soft band (0.5 - 2 keV) for the XDF \citep{Luo+17} and obtain $L_{\rm 2keV} \approx 3.48 \times 10^{25}\rm~erg/s/Hz$. 
These estimates imply $\alpha_{\rm ox} \gtrsim 1.25$, a range of values for which the rest-frame UV emission of the galaxy dominates over the X-rays, thus confirming the intrinsic weak X-ray nature of {\it Virgil}.

In addition to the X-ray-based catalogs of AGN, we searched for possible counterparts for {\it Virgil} in the catalog by \citet{Lyu+22}, which extended AGN detection in the HUDF by analysing galaxies' rest-frame optical to mid-infrared emission. We also examined the recently released AGN catalog by \citet{Lyu+24}, based solely on MIRI imaging from the SMILES program. 
In both cases, no previously identified AGN is found at the coordinates of our target.

\begin{figure}
    \centering
    \includegraphics[width = .5\textwidth]{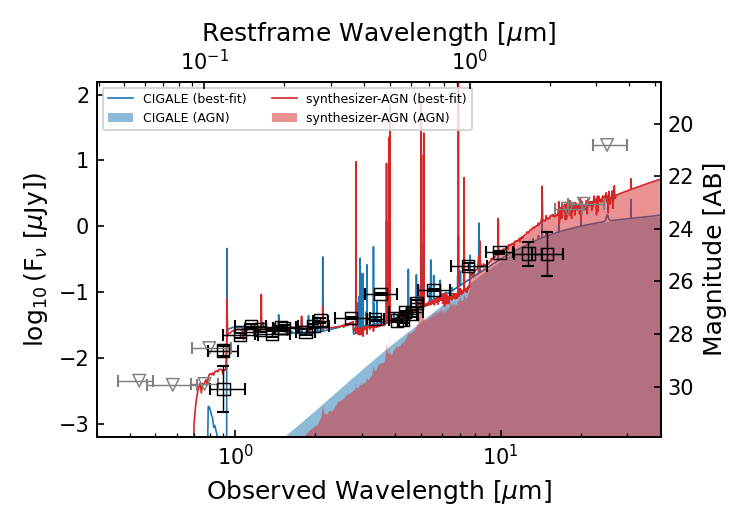}
    \caption{Decomposition of the AGN component from {\it Virgil}'s photometry (black open squares and gray triangles for the upper limits) for the best-fit models obtained with the SED fitting codes {\sc CIGALE} (blue) and {\sc synthesizer-AGN} (red).}
    \label{fig:sed_agn_component}
\end{figure}

As a result of our SED-fitting runs on the whole available photometry (ALMA upper limits included), we find that all codes prefer the SED solution implementing an AGN to the case of only stellar populations, delivering better quality fits (i.e. lower $\chi^2_{\rm red}$),  
see Figure~\ref{fig:virgil_sed} (bottom panel) and Table~\ref{tab:sed-agn}.
In this case, the best-fits are provided by {\sc synthesizer-AGN} ($\chi2_{\rm red} = 5.4$) and {\sc CIGALE} ($\chi2_{\rm red} = 6.9$).  
For {\sc synthesizer-AGN}, the best-fit indicates a total stellar mass $\rm M_\star \simeq 10^{8.2}~M_\odot$ with two young stellar populations: a 1~Myr old stellar population with a metallicity $Z = 0.2~\rm Z_\odot$ and $A_V = 0.5$, and a 25 Myr old stellar population of solar metallicity and $A_V = 0.9$.
The smaller stellar mass of this run with respect to the no-AGN case is due to the dominant contribution at the NIR wavelengths of the AGN component.
In the case of {\sc CIGALE}, {\it Virgil}'s best-fit has a stellar mass of $\rm M_\star \simeq 10^{8.5 \pm 0.2}~M_\odot$ and is reproduced with a young (3~Myr) and an older (300~Myr) stellar population with solar metallicities and $A_V = 0.8$.
Also for {\sc CIGALE}, the AGN dominates at the longest MIRI wavelengths 
, see Figure~\ref{fig:sed_agn_component}.

We further investigate the properties of the eventual AGN hosted in {\it Virgil} by estimating its bolometric luminosity $L_{\rm AGN,bol}$ and supermassive black-hole mass $M_{\rm BH}$.
Following \cite{Kaspi+00}, we derive $L_{\rm AGN,bol}$ from the monochromatic luminosity at 5100\AA\ (corrected for dust extinction) $L$(5100\AA) as $L_{\rm AGN, bol} \approx 9 \cdot \lambda \cdot L$(5100\AA).
In the case of {\it Virgil}, the estimated luminosity at 5100\AA\ is $L(\rm 5100\AA) \approx 2.9 \times 10^{39}\rm~ erg~s^{-1}~\AA^{-1}$ . However, according to our best-fit with CIGALE, at 5100\AA\ (rest-frame) the AGN contributes only to 40 percent of the observed emission. 
Besides, the AGN's dust attenuation is $A(5100{\rm \AA}) \approx 3.3\rm~mag$. Therefore, correcting for AGN contribution and dust obscuration, we estimate $L_{\rm AGN}({\rm 5100\AA}) \approx 2.4 \times 10^{40}\rm~ erg~s^{-1}~\AA^{-1}$ that converts into an $L_{\rm AGN,bol} \approx 1.1 \times 10^{45}\rm~ erg~s^{-1}$.
We recover similar values also when considering the best-fit results from \textsc{synthesizer-AGN}: $L_{\rm AGN}({\rm 5100\AA}) \approx 1.9 \times 10^{40}\rm~ erg~s^{-1}~\AA^{-1}$ and $L_{\rm AGN,bol} \approx 8.9 \times 10^{44}\rm~ erg~s^{-1}$. 
By comparing the so-derived AGN bolometric luminosity to the Eddington luminosity $L_{\rm Edd}$ \citep{Eddington+26}, we can infer $M_{\rm BH}$ via the relation $M_{\rm BH} = 7.9 \times 10^{-39} \times (L_{\rm AGN,bol} / \lambda_{\rm Edd})$, where $\lambda_{\rm Edd}$ is the Eddington ratio.
Considering typical values for the Eddington ratio $\lambda_{\rm Edd} = 0.1 - 1$, the range of values we obtain for the mass of the supermassive black hole of our target varies from $M_{\rm BH} \approx (7 - 9)\times 10^6\rm~ M_\odot$ ($\lambda_{\rm Edd} = 1$) to $M_{\rm BH} \approx (7 - 9) \times 10^7\rm~ M_\odot$ ($\lambda_{\rm Edd} = 0.1$).
These estimates are in good agreement with the values of bolometric luminosity of AGNs at $z > 4$ recently studied with JWST \citep{Greene+24, Furtak+24, Kokorev+23, Matthee+24}.

\begin{deluxetable*}{lcccccccc}[t]
    \tablenum{2}
    \tablecaption{Stellar SED properties of \textit{Virgil} - only stars}
    \tablewidth{0pt}
    \tablehead{
\colhead{Code}          & \colhead{$\chi^2_{\rm red}$} & \colhead{$\log_{10}(\rm M_\star)$}   & \colhead{$\rm A_{V, main}$} & \colhead{Age$_{\rm main}$}   & \colhead{$\rm A_{V, young}$} & \colhead{Age$_{\rm young}$}   & \colhead{Age$_{\rm mass-weighted}$} & \colhead{SFR$_{\rm 100Myr}$}          \\  
                         &                              & \colhead{$\log_{10}(\rm [M_\odot])$} & \colhead{[mag]}             & \colhead{[Myr]}              & \colhead{[mag]}              & \colhead{[Myr]}               & \colhead{[Myr]}                     & \colhead{[M$_\odot~\rm yr^{-1}$]} }  
                         
\startdata
\textsc{synthesizer-AGN} &  6.5 &  $9.71 \pm 0.10$                & $4.00 \pm 0.1$                & $26 \pm 4$               & $0.50 \pm 0.01$ & $6 \pm 1$            & $32^{+9}_{-7}$                   & -                      \\
\textsc{cigale}          & 19.4 &  $8.95 \pm 0.08$        & $0.81 \pm 0.04$        & $304 \pm 223$       & -       & $3 \pm 1$       & $275 \pm 104$       & $2.60 \pm 0.87$        \\
\textsc{bagpipes}        & 20.6 &  $8.74^{+0.38}_{-0.12}$ & $0.73^{+0.05}_{-0.09}$ & $720^{+70}_{-230}$  & -       & $20^{+2}_{-1}$  & $220^{+370}_{-130}$ & $3.42^{+0.67}_{-1.35}$ \\ 
\textsc{prospector}      & 25.1 &  $9.01^{+0.03}_{-0.16}$ & $0.12^{+0.02}_{-0.01}$ &   -                 & -       & -               & $367^{+51}_{-33}$               & $2.53^{+0.51}_{-1.90}$              \\                 
\enddata
    \tablecomments{In the above table, we report the main physical properties of {\it Virgil} as derived from the different SED fitting codes adopted in our study: \textsc{synthesizer-AGN}, \textsc{cigale}, \textsc{bagpipes}, and \textsc{prospector}.
    In this table, we present the results derived when considering only the contribution of stellar populations to the overall {\it Virgil}'s SED. Specifically, we report the logarithm of the stellar mass $\log_{10}(\rm M_\star)$, the extinction of the main stellar component $\rm A_{V,main}$ and its age Age$_{\rm main}$, the extinction of the young stellar component $\rm A_{V,young}$ and its age Age$_{\rm young}$, the mass-weighted age of the overall galaxy $\rm Age_{mass-weighted}$ and its average star formation rate over the last 100 Myr SFR$_{\rm 100Myr}$. We leave an empty entry for parameters where the codes do not provide an estimate.}
    \label{tab:sed-no-agn}
\end{deluxetable*}

\begin{deluxetable*}{lcccccccc}[t]
    \tablenum{3}
    \tablecaption{Stellar SED properties of \textit{Virgil} - stars \& AGN}
    \tablewidth{0pt}
    \tablehead{
\colhead{Code}          & \colhead{$\chi^2_{\rm red}$} & \colhead{$\log_{10}(\rm M_\star)$}   & \colhead{$\rm A_{V, main}$} & \colhead{Age$_{\rm main}$}   & \colhead{$\rm A_{V, young}$} & \colhead{Age$_{\rm young}$}   & \colhead{Age$_{\rm mass-weighted}$} & \colhead{SFR$_{\rm 100Myr}$}          \\  
                         &                              & \colhead{$\log_{10}(\rm [M_\odot])$} & \colhead{[mag]}             & \colhead{[Myr]}              & \colhead{[mag]}              & \colhead{[Myr]}               & \colhead{[Myr]}                     & \colhead{[M$_\odot~\rm yr^{-1}$]}     
              }  
\startdata
{\sc synthesizer-AGN} &  5.4 &  $8.19 \pm 0.07$        & $0.90\pm 0.10$         & $2.5 \pm 0.5$      & $0.46 \pm 0.03$ & $1 \pm 0.5$       & $0.8^{+0.8}_{-0.2}$ &  -                     \\                                     
{\sc cigale}          &  6.9 &  $8.50 \pm 0.02$        & $0.83 \pm 0.04$        & $662 \pm 230$      & -               & $1.5 \pm 0.5$     & $309 \pm 121$       & $0.88 \pm 0.04$        \\                                         
{\sc bagpipes}        & 12.1 & $10.25^{+0.04}_{-0.05}$ & $3.57^{+0.18}_{-0.17}$ & $790^{+20}_{-30}$  & -               & $260^{+20}_{-60}$ & $640^{+20}_{-110}$  &  -                     \\ 
{\sc prospector}      & 23.4 &  $8.77^{+0.06}_{-0.11}$ & $0.11^{+0.01}_{-0.02}$ &    -               & -               &  -                & $284^{+31}_{-32}$  & $2.50^{+0.53}_{-1.75}$ \\                           
\enddata
    \tablecomments{As in Table~\ref{tab:sed-no-agn}, we report the main physical properties of {\it Virgil} as derived from the different SED fitting codes adopted in our study.
    In this table, we present the results derived when adding to the contribution of stellar populations also an AGN component.}
    \label{tab:sed-agn}
\end{deluxetable*}

\subsubsection{{\it Virgil}, an LRD?}
\label{sec:lrd_like}
The flat rest-frame UV and the red NIR colors of {\it Virgil} revealed by our multi-wavelength dataset (HST, JWST NIRCam and MIRI) are similar to the ``v-shaped'' SED (in the $\lambda - f_{\lambda}$ plane) of the recently discovered population of LRDs \cite[e.g.][]{Labbe+23b, Furtak+23, Kokorev+24, Perez-Gonzalez+24b, Akins+24, Matthee+24}.
Despite the rapid emergence of LRDs in the literature over the past two years, their nature and unusual SEDs remain largely unexplained. Proposed explanations suggest a complex interplay between AGN activity, stellar populations, and dust \citep{Labbe+23b, Li+24, Inayoshi+24}. 
While this debate continues, the detection of broad Balmer lines (e.g., H$\alpha$, H$\beta$) in some LRDs has been interpreted as strong evidence for AGN activity, implying accretion onto supermassive black holes \citep{Furtak+23b, Fujimoto+23, Greene+24, Killi+23, Kokorev+23, Matthee+24}. 

Following \cite{Kokorev+24}, {\it Virgil}'s photometry complies with most of the color criteria used to characterise LRDs at $z > 6$. In particular, for {\it Virgil} we find: 

\begin{itemize}
\item F150W - F200W~$ = 0.05 \pm 0.10~ (< 0.8)$;
\item F277W - F356W~$ = 0.92 \pm 0.07~  (> 0.6)$; 
\item F277W - F444W~$ = 0.22 \pm 0.07~ (> 0.7)$;
\item F115W - F200W~$ = -0.01\pm 0.10~ (> -0.5)$;
\end{itemize}

where the values in parentheses indicate the selection criteria from \citet{Kokorev+24}, designed to identify ``v-shaped'' SEDs while minimizing contamination from brown dwarfs (particularly the F115W - F200W condition).
The only discrepancy is {\it Virgil}'s F277W - F444W color, which appears bluer than expected for LRDs. However, at the specific redshift of our target ($z = 6.63124$), this could be explained by the positioning of the filters relative to strong nebular lines. 
The F444W filter probes wavelengths redward of [\ion{O}{3}] and blueward of H$\alpha$.
At the wavelengths of these strong optical emission lines, the transmission drops below 5\%, making the F444W quite robustly probe the optical continuum without significant contamination from such strong transitions. 
This may result in a bluer color. 
Indeed, replacing F444W with the adjacent MIRI filter F560W reveals a significantly redder color: F277W - F560W~$= 1.10 \pm 0.06$.

At rest-frame UV wavelengths, {\it Virgil} meets the brown dwarf rejection criteria from \citet{Kokorev+24}. 
However, it does not strictly comply with the F150W - F200W color threshold proposed by \citet{Langeroodi2023c}, who suggests that objects with F150W - F200W ~$< 0.25$ are typically brown dwarfs. 
Nonetheless, {\it Virgil}'s F277W - F444W color places it outside the region occupied by brown dwarfs in the F277W - F444W vs. F150W - F200W diagram (see Figure 13 of \citealt{Langeroodi2023c}). 
Additionally, its extended morphology further rules out the brown dwarf hypothesis.

The estimate of the compactness parameter ($c = f(r=0\farcs2)/f(r=0\farcs1)$ with fluxes properly corrected for aperture) derived from the F444W image after removing contamination from the nearby LAE (see \S\ref{sec:decontamination}) yields $c_{\rm F444W} \approx 1.5$, thus satisfying the LRD compactness criterion $c_{\rm F444W} < 1.7$.

Despite not fully meeting the LRDs' F277W - F444W color, {\it Virgil} complies with all the other LRD criteria adopted in the recent literature. This shows that our target is likely to be the first LRD with a detectable host galaxy from rest-frame UV-to-optical wavelengths (see \S~\ref{sec:morphology}). 
In this regard, \citet{Killi+23} already reported an LRD at $z \approx 4.5$ with a slightly more extended rest-frame UV morphology than previously assumed for this class of sources.
Furthermore, following our discovery of {\it Virgil}, additional LRDs have been identified with extended and more complex UV morphologies than simple PSF-like profiles \citep{Rinaldi+25}. 
{\it Virgil}, however, would be the first LRD with a clear detection of its host galaxy, having a resolved morphology down to its rest-frame optical wavelengths.
Overall, these findings suggest a broader diversity among LRDs than initially anticipated.

\begin{figure}
    \centering
    \includegraphics[width = .5\textwidth]{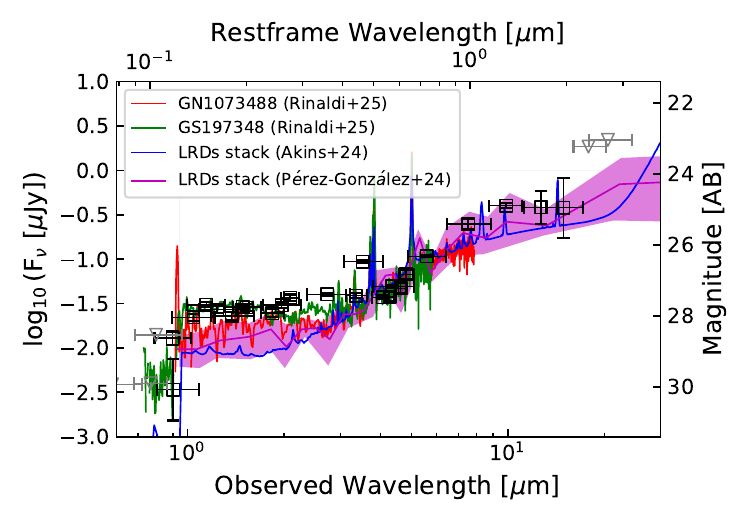}
    \caption{{\it Virgil}'s photometry (black open squares and gray open triangles for the $3\sigma$ upper limits) and the average observed SED of MIRI-detected LRDs (in magenta) found in the HUDF and reported by \cite{Perez-Gonzalez+24a}. 
    In blue the ``maximal'' SED of 500 LRDs detected in COSMOS-Webb \citep{Akins+24} while in red and green the NIRCam/MSA spectra of two LRDs in HUDF \cite[][]{Rinaldi+25}.} 
    \label{fig:empirical_models}
\end{figure}

To further corroborate {\it Virgil}'s resemblance to LRDs, we compare its photometry to empirical SEDs derived from stacked LRDs, including the average SED of 20 MIRI-detected LRDs in the HUDF \citep{Perez-Gonzalez+24a} and the ``maximal'' SED from stacking 500 LRDs in the COSMOS-Webb field \citep{Akins+24} (see Figure~\ref{fig:empirical_models}).
After redshifting these empirical SEDs to $z = 6.63124$ and normalizing their fluxes to match {\it Virgil}'s photometry, we find a strong agreement at rest-frame optical/NIR wavelengths. 
However, at shorter wavelengths (rest-frame UV), the empirical SEDs lie below {\it Virgil}'s observed photometry. 
This could be the wavelength regime where {\it Virgil}'s underlying stellar emission could contribute more to its overall emission, suggesting that its stronger UV emission could be mainly attributed to young stellar populations as also found from our SED fitting (see \S~\ref{sec:sed}).

We also compare {\it Virgil}'s photometry with the NIRCam/MSA spectra of two LRDs detected in the HUDF \citep{Rinaldi+25}. 
The best match is with GS197348 \citep[green spectrum;][]{Bunker+23, Rinaldi+25}, which closely reproduces {\it Virgil}’s SED, even at rest-frame UV wavelengths. 
This further highlights that while LRDs share similar optical/NIR properties, their UV emission can vary significantly from object to object.

Interestingly, {\it Virgil}'s non-detection in X-rays (\S\ref{sec:full_sed}) is consistent with recent findings for LRDs \citep{Yue+24, Ananna+24}.
Similarly, also the non-detection at far-infrared wavelengths (ALMA upper limits) is in agreement with general LRD findings \citep{Labbe+23b, Perez-Gonzalez+24a, Williams+24}.

Once again, we emphasize the critical role of MIRI data in this classification. 
Without MIRI imaging, identifying {\it Virgil} as a ``v-shaped'' object would have been impossible.
In addition, we highlight that existing LRD selection criteria (based solely on NIRCam photometry) extend only to rest-frame optical wavelengths near H$\alpha$ at $z > 6$. 
This means that they miss the rising NIR red continuum of LRDs and are susceptible to contamination from strong line-emitters \cite[e.g., ][]{Hainline+25}. 
Incorporating MIRI data is therefore essential for robustly identifying this population at $z > 6$.

\subsection{The impact of long-wavelengths MIRI filters}
A final remark comes from the information that we can obtain thanks to the presence of the longest MIRI wavelengths (i.e. F1280W, F1500W).
Despite the shallower depth of the available imaging above 10~$\mu$m, we find that the steep rise observed up to F1000W could tend to flatten at longer wavelengths (above the rest-frame 1.6~$\mu$m).
Interestingly, such flattening of the SED was also reported in the average trend of the MIRI-detected LRDs by \cite{Williams+24} and \cite{Perez-Gonzalez+24a}.
Also \cite{Wang+24} revealed an unexpected flattening at rest-frame wavelengths 0.7 - 5 $\mu$m in their spectro-photometric study of an LRD at $z \approx 3.1$.
The origin of this flattening is still debated but, in the AGN scenario, one possible explanation \citep{Wang+24} could be a lack of torus emission as in the case of hot-dust-deficient AGNs \cite[e.g., ][]{Jiang+10, Lyu+17, Son+23}.

Due to the higher redshift of our target, we have a hint of flattening only from the F1280W and F1500W filters, probing the rest-frame 1.6 - 2 $\mu$m wavelength range. 
We retrieve an estimate of the flux in F1280W (S/N~$\approx 2.4$) and F1500W (S/N~$\approx 1.3$) that imply colors of F1000W - F1280W~$= 0.1 \pm 0.2$ and F1000W - F1500W~$ = 0.1 \pm 0.3$.
The shallower depth of the available MIRI filters $\lambda > 18~\mu$m prevents us from robustly confirming the flattening on a longer wavelength range. 
Deeper F1800W and F2100W imaging would be necessary to extend this finding up to $3 \mu$m (rest-frame).

\begin{figure}
    \centering
    \includegraphics[width = .5\textwidth]{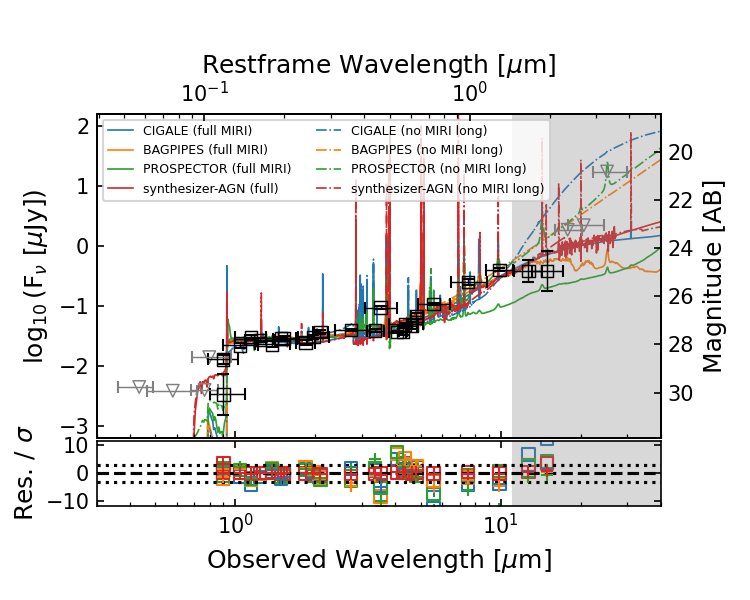}
    \caption{{\it Virgil}'s photometry (as in the previous panels) and best-fit models obtained with the SED fitting codes {\sc CIGALE} (blue), {\sc Bagpipes} (orange), {\sc prospector} (green) and  {\sc synthesyzer-AGN} (red)
    in the case of excluding (dashed-dotted lines) and assuming (solid lines) the photometric MIRI information at $> 10 \mu$m (gray-shaded area). Similarly, the bottom panel shows the ratio between the residuals and the photometric error for all the available bands when excluding (plus) and assuming (open squares) the MIRI long wavelength filters. The dotted lines are indicative of residuals equal to $\pm 3\sigma$.}
    \label{fig:sedmiri_short}
\end{figure}

We underline how, if the flattening trend is confirmed, without any information at these wavelengths the SED codes would tend to prefer models that keep on rising thus over-predicting the flux at the longest wavelengths and overestimating the importance of the AGN dust torus, see Figure~\ref{fig:sedmiri_short}.
This further underlines the importance of conducting deep MIRI imaging above the 10~$\mu$m even if challenged by the less sensitivity of the MIRI instrument at such wavelengths.

\section{Summary and Conclusions}
\label{sec:conclusions}

In this paper, we have presented the analysis of {\it Virgil}, a MIRI extremely red object (MERO) detected with the F1000W filter as 
part of the MIRI Deep Imaging Survey (MIDIS) observations of the Hubble Ultra Deep Field.
{\it Virgil} is a Ly$\alpha$ emitter at $z = 6.63124 \pm 0.00188$ (VLT/MUSE, \citealt{Bacon+23}) showing 
a red F444W - F1000W~$= 2.33\pm0.06$ color.

At NIRCam wavelengths, {\it Virgil} presents an elongated morphology in the south-west direction and features a tail that ends up in a (faint) compact source ({\it Beatrix}; $\alpha$(J2000.0) = 03:32:37.8844 (hours); $\delta$(J2000.0) = -27:47:10.979). With the current data, it is, however, difficult to assess if this compact object is a clump of {\it Virgil}, a clump of a nearby ($\delta$ = 0\farcs4 - 0\farcs7) foreground ($z \simeq 4.77$) LAE \citep{Matthee+22} , or a different source.  
By modeling the rest-frame UV-optical morphology of {\it Virgil} with a S\'ersic profile we find a best-fit S\'ersic index $n = 0.93^{+0.85}_{-0.31}$ and an effective radius $r_e = 0.49^{+0.05}_{-0.11}$~pkpc in agreement with the values recently reported for other galaxies at similar redshifts \citep{Sun+24}.
At MIRI wavelengths $> 7\mu$m, the morphology of {\it Virgil} is dominated by the MIRI PSF.

The extracted photometry of {\it Virgil} shows a blue UV continuum ($\beta_{\rm UV} = -2.1 \pm 0.3$),  the presence of strong emission lines (flux excess of the F356W band corresponding to the H$\beta$ + [\ion{O}{3}] complex) and a steep rising SED from 0.7 to 1.3 $\mu$m (rest-frame). 
SED fitting models considering only stellar populations struggle to explain {\it Virgil}'s behaviour at MIRI wavelengths.
The best-fit model suggest that {\it Virgil}' SED originates from two components: an extremely young (6 Myr) and unobscured stellar population dominating the blue wavelengths, and another young (26 Myr) stellar population strongly affected by extinction (A$_V = 4$~mag) that is responsible for most of the light at red wavelengths. 
In agreement with recent papers on LRDs \citep{Williams+24, Perez-Gonzalez+24a, Barro+23}, in this scenario the bulk of the dust re-emitting the absorbed light at rest-frame UV and optical wavelengths should be hotter than 60 - 70~K to comply with the ALMA upper limits. 
This temperature is significantly higher than the one typically adopted for SED fitting \cite[i.e. 20 - 30~K; e.g., ][]{Chary+01}.
In this case, {\it Virgil} could be similar to a dusty starburst.
The small angular size of {\it Virgil} together with the coarser MIRI resolution hamper us from deriving if the two populations are located in different parts of the galaxy.  
All the other codes fail to reproduce the steep and rising MIRI photometry.

Such red rising part of {\it Virgil}'s SED could hint at the presence of a (dust-obscured) AGN.
Despite extensive searches, no AGN counterparts are found in existing multi-wavelength catalogs \citep{Ranalli+13, Luo+17, Evans+20, Lyu+22, Lyu+24}. 
However, when implementing an AGN contribution to the SED, all the SED-fitting codes adopted deliver significantly improved best-fit models.
Assuming the presence of an AGN, we estimate a bolometric luminosity $L_{\rm bol, AGN} \approx (8.9 - 11) \times 10^{44}\rm~ erg~s^{-1}$, and an $M_{\rm BH} = (7 - 9)\times 10^6\rm~M_\odot$ (if $\lambda_{\rm Edd} = 1$).
These estimates agree with those reported in the recent literature targeting reddened AGNs at $z > 4$ \citep[e.g.,][]{Kokorev+23, Scholtz+23, Matthee+24, Furtak+24, Rinaldi+25}.
Nonetheless, with the currently available data is difficult to fully rule out one of the proposed scenarios. We cannot even exclude that the two conditions (dusty starburst and AGN) could be true at the same time, i.e. not too differently from what was recently reported for the GN20 galaxy at $z = 4.05$ \cite[e.g.,][]{Colina+23, Bik+24, Crespo+24, Ubler+23}. 

{\it Virgil}'s flat rest-frame UV and red NIR colors and, in general, the shape of its SED, suggest that our target belongs to the recently discovered population of LRDs \cite[e.g.][]{Labbe+23b, Kokorev+24, Perez-Gonzalez+24a}. 
This is further confirmed by the comparison of {\it Virgil}' SED with the recently reported empirical SED \citep{Perez-Gonzalez+24a, Akins+24} and the actual spectra of other LRDs \citep{Rinaldi+25}. 
However, the extended morphology of {\it Virgil} at UV-to-optical wavelengths makes it the first LRD for which we can clearly detect the host galaxy.

Despite the richness of the photometric dataset available, deeper MIRI long-wavelength observations in combination with a rest-frame optical and near-infrared spectroscopic follow-up \cite[as demonstrated in recent literature, e.g.][]{Wang+24, Tacchella+24, Li+24} appear to be fundamental to properly assess the nature of {\it Virgil}.


All in all, our findings show the power of MIRI imaging in unveiling the complexity of this galaxy's nature.
For our target, MIRI provides a fundamental piece of information that would be otherwise completely missed without imaging at such long wavelengths. 
This discovery opens up to the question of how many objects like {\it Virgil} there are in the Universe.
If a systematic search for similar objects in deep MIRI surveys would reveal several {\it Virgil}-like objects and their AGN-nature is confirmed, this discovery could potentially bring us to reassess (once more) the role of AGN to the reionization.
In addition to a more general importance of MIRI in the characterisation of galaxy properties, this study also advocates for the importance of conducting deep and extended MIRI surveys at $\lambda > 10~\mu$m wavelengths, 
a regime that results in being crucial in correctly determining the physical properties of galaxies and unveiling their hidden components.
In fact, the absence of MIRI long-wavelength photometry (even in the case of upper limits) biases us to overestimate the contribution of AGN in the SED of such objects.

\section*{Acknowledgments}
The authors thank R. Cooper, G. Yang, V. Kokorev, D. Wen, C. Williams, H. \"Ubler for useful discussions and comments.

E.I. and K.I.C. acknowledge funding from the Netherlands Research School for Astronomy (NOVA). 
K.I.C. acknowledges funding from the Dutch Research Council (NWO) through the award of the Vici Grant VI.C.212.036.
A.A.-H. acknowledges support from grant PID2021-124665NB-I00 funded by MCIN/AEI/10.13039/ 501100011033 and by “ERDF A way of making Europe”.
P.G.P-G. acknowledges support from grant PID2022-139567NB-I00 funded by Spanish Ministerio de Ciencia e Innovaci\'on MCIN/AEI/10.13039/501100011033,
FEDER {\it Una manera de hacer
Europa}.
J.A.-M., A.C.-G. and L. Colina,
acknowledge support by grant PIB2021-127718NB-100 from the Spanish Ministry of Science and Innovation/State Agency of Research MCIN/AEI/10.13039/501100011033 and by “ERDF A way of making Europe”.
L. Colina thanks the support from the Cosmic Dawn Center received during visits to DAWN as international associate.
L. Costantin acknowledges support by grants PIB2021-127718NB-100 and PID2022-139567NB-I00 from the Spanish Ministry of Science and Innovation/State Agency of Research MCIN/AEI/10.13039/501100011033 and by “ERDF A way of making Europe”.
T.R.G acknowledges support from the Carlsberg Foundation (grant no CF20-0534).
S.G. acknowledges financial support from the Cosmic Dawn Center (DAWN), funded by the Danish National Research Foundation (DNRF) under grant No. 140.
This work was supported by research grants (VIL16599, VIL54489) from VILLUM FONDEN.
J.P.P. and T.V.T. acknowledge financial support from the UK Science and Technology Facilities Council, and the UK Space Agency.

This work is based on observations made with the NASA/ESA/CSA James Webb Space Telescope. The data were obtained from the Mikulski Archive for Space Telescopes at the Space Telescope Science Institute, which is operated by the Association of Universities for Research in Astronomy, Inc., under NASA contract NAS 5-03127 for JWST. These observations are associated with programs GO \#1963, GO \#1895 and GTO \#1283. 
The authors acknowledge the team led by coPIs C. Williams, M. Maseda and S. Tacchella, and PI P. Oesch, for developing their respective observing programs with a zero-exclusive-access period. 
Also based on observations made with the NASA/ESA Hubble Space Telescope obtained from the Space Telescope Science Institute, which is operated by the Association of Universities for Research in Astronomy, Inc., under NASA contract NAS 5–26555.  
The work presented here is the effort of the entire MIRI team and the enthusiasm within the MIRI partnership is a significant factor in its success. MIRI draws on the scientific and technical expertise of the following organisations: Ames Research Center, USA; Airbus Defence and Space, UK; CEA-Irfu, Saclay, France; Centre Spatial de Li\`ege, Belgium; Consejo Superior de Investigaciones Cient\'{\i}ficas, Spain; Carl Zeiss Optronics, Germany; Chalmers University of Technology, Sweden; Danish Space Research Institute, Denmark; Dublin Institute for Advanced Studies, Ireland; European Space Agency, Netherlands; ETCA, Belgium; ETH Zurich, Switzerland; Goddard Space Flight Center, USA; Institute d’Astrophysique Spatiale, France; Instituto Nacional de T\'ecnica Aeroespacial, Spain; Institute for Astronomy, Edinburgh, UK; Jet Propulsion Laboratory, USA; Laboratoire d’Astrophysique de Marseille (LAM), France; Leiden University, Netherlands; Lockheed Advanced Technology Center (USA); NOVA Opt-IR group at Dwingeloo, Netherlands; Northrop Grumman, USA; Max-Planck Institut für Astronomie (MPIA), Heidelberg, Germany; Laboratoire d’Etudes Spatiales et d’Instrumentation en Astrophysique (LESIA), France; Paul Scherrer Institut, Switzerland; Raytheon Vision Systems, USA; RUAG Aerospace, Switzerland; Rutherford Appleton Laboratory (RAL Space), UK; Space Telescope Science Institute, USA; Toegepast-Natuurwetenschappelijk Onderzoek (TNO-TPD), Netherlands; UK Astronomy Technology Centre, UK; University College London, UK; University of Amsterdam, Netherlands; University of Arizona, USA; University of Cardiff, UK; University of Cologne, Germany; University of Ghent; University of Groningen, Netherlands; University of Leicester, UK; University of Leuven, Belgium; University of Stockholm, Sweden; Utah State University, USA.

For the purpose of open access, the author has applied a Creative Commons Attribution (CC BY) licence to the Author Accepted Manuscript version arising from this submission.

\vspace{5mm}
\facilities{{\sl VLT}, {\sl HST}, {\sl JWST}, {\sl ALMA}}.

\software{\textsc{Astropy} \citep{astropy:2013, astropy:2018, astropy:2022}, 
\textsc{NumPy} \citep{Harris+20},
\textsc{pandas} \citep{McKinney+10, reback2020pandas}
\textsc{Photutils} \citep{Bradley+22}, 
\textsc{SciPy} \citep{Virtanen+20},
\textsc{WebbPSF} \citep{Perrin+14},
\textsc{Source Extractor} \citep{Bertin+96},
\textsc{lenstronomy}
\citep{Birrer+18},
\textsc{sep}
\citep{Barbary+17},
\textsc{LePHARE} \citep{Ilbert+06, LePHARE}, 
\textsc{Eazy}
\citep{Brammer+08},
\textsc{Bagpipes} \citep{Carnall+18},
\textsc{CIGALE}
\citep{Burgarella+05,Noll+09,Boquien+19},
\textsc{prospector}
\citep{Johnson+21},
\textsc{synthesizer-AGN}
\citep{Perez-Gonzalez+03, Perez-Gonzalez+08},
\textsc{AstroPhot}
\citep{Stone+23},
\textsc{TOPCAT} \citep{Taylor+11}.
          }


\bibliography{Virgil}{}
\bibliographystyle{aasjournal}

\end{document}